\documentclass[prd,superscriptaddress,unsortedaddress,twocolumn,showpacs,preprintnumbers,amsmath,amssymb]{revtex4}
\usepackage[dvips]{graphicx}
\usepackage{amsmath,amssymb}
\usepackage{ulem}

\newcommand{\beq}{\begin{equation}}
\newcommand{\eeq}{\end{equation}}
\newcommand{\bea}{\begin{eqnarray}}
\newcommand{\eea}{\end{eqnarray}}

\DeclareSymbolFont{boldletters}{OML}{cmm} {b}{it}
\DeclareSymbolFontAlphabet{\mathbit}{boldletters}
\DeclareMathSymbol{\alpha}{\mathalpha}{letters}{"0B}
\DeclareMathSymbol{\beta}{\mathalpha}{letters}{"0C}
\DeclareMathSymbol{\gamma}{\mathalpha}{letters}{"0D}
\DeclareMathSymbol{\delta}{\mathalpha}{letters}{"0E}
\DeclareMathSymbol{\epsilon}{\mathalpha}{letters}{"0F}
\DeclareMathSymbol{\zeta}{\mathalpha}{letters}{"10}
\DeclareMathSymbol{\eta}{\mathalpha}{letters}{"11}
\DeclareMathSymbol{\theta}{\mathalpha}{letters}{"12}
\DeclareMathSymbol{\iota}{\mathalpha}{letters}{"13}
\DeclareMathSymbol{\kappa}{\mathalpha}{letters}{"14}
\DeclareMathSymbol{\lambda}{\mathalpha}{letters}{"15}
\DeclareMathSymbol{\mu}{\mathalpha}{letters}{"16}
\DeclareMathSymbol{\nu}{\mathalpha}{letters}{"17}
\DeclareMathSymbol{\xi}{\mathalpha}{letters}{"18}
\DeclareMathSymbol{\pi}{\mathalpha}{letters}{"19}
\DeclareMathSymbol{\rho}{\mathalpha}{letters}{"1A}
\DeclareMathSymbol{\sigma}{\mathalpha}{letters}{"1B}
\DeclareMathSymbol{\tau}{\mathalpha}{letters}{"1C}
\DeclareMathSymbol{\upsilon}{\mathalpha}{letters}{"1D}
\DeclareMathSymbol{\phi}{\mathalpha}{letters}{"1E}
\DeclareMathSymbol{\chi}{\mathalpha}{letters}{"1F}
\DeclareMathSymbol{\psi}{\mathalpha}{letters}{"20}
\DeclareMathSymbol{\omega}{\mathalpha}{letters}{"21}
\DeclareMathSymbol{\varepsilon}{\mathalpha}{letters}{"22}
\DeclareMathSymbol{\vartheta}{\mathalpha}{letters}{"23}
\DeclareMathSymbol{\varpi}{\mathalpha}{letters}{"24}
\DeclareMathSymbol{\varrho}{\mathalpha}{letters}{"25}
\DeclareMathSymbol{\varsigma}{\mathalpha}{letters}{"26}
\DeclareMathSymbol{\varphi}{\mathalpha}{letters}{"27}
\DeclareMathSymbol{\Gamma}{\mathalpha}{letters}{"00}
\DeclareMathSymbol{\Delta}{\mathalpha}{letters}{"01}
\DeclareMathSymbol{\Theta}{\mathalpha}{letters}{"02}
\DeclareMathSymbol{\Lambda}{\mathalpha}{letters}{"03}
\DeclareMathSymbol{\Xi}{\mathalpha}{letters}{"04}
\DeclareMathSymbol{\Pi}{\mathalpha}{letters}{"05}
\DeclareMathSymbol{\Sigma}{\mathalpha}{letters}{"06}
\DeclareMathSymbol{\Upsilon}{\mathalpha}{letters}{"07}
\DeclareMathSymbol{\Phi}{\mathalpha}{letters}{"08}
\DeclareMathSymbol{\Psi}{\mathalpha}{letters}{"09}
\DeclareMathSymbol{\Omega}{\mathalpha}{letters}{"0A}

\begin{document}
\preprint{SAGA-HE-239-08}
\title{
Phase diagram in the imaginary chemical potential region and 
extended ${\mathbb Z}_3$ symmetry}

\author{Yuji Sakai}
\email[]{sakai2scp@mbox.nc.kyushu-u.ac.jp}
\affiliation{Department of Physics, Graduate School of Sciences, Kyushu University,
             Fukuoka 812-8581, Japan}
\author{Kouji Kashiwa}
\email[]{kashiwa@phys.kyushu-u.ac.jp}
\affiliation{Department of Physics, Graduate School of Sciences, Kyushu University,
             Fukuoka 812-8581, Japan}

\author{Hiroaki Kouno}
\email[]{kounoh@cc.saga-u.ac.jp}
\affiliation{Department of Physics, Saga University,
             Saga 840-8502, Japan}

\author{Masanobu Yahiro}
\email[]{yahiro@phys.kyushu-u.ac.jp}
\affiliation{Department of Physics, Graduate School of Sciences, Kyushu University,
             Fukuoka 812-8581, Japan}

\date{\today}

\begin{abstract}
Phase transitions in the imaginary chemical potential region 
are studied by the Polyakov loop extended Nambu--Jona-Lasinio 
(PNJL) model that  possesses the extended ${\mathbb Z}_{3}$ symmetry. 
The extended-${\mathbb Z}_{3}$ 
invariant quantities such as the partition function, 
the chiral condensate and the modified Polyakov loop 
have the Roberge-Weiss (RW) periodicity. 
There appear four types of phase transitions; 
deconfinement, chiral, Polyakov-loop RW and chiral RW transitions. 
The orders of the chiral and deconfinement transitions depend 
on the presence or absence of current quark mass, 
but those of the Polyakov-loop RW and chiral RW transitions do not. 
The scalar-type eight-quark interaction newly added in the model makes 
the chiral transition line shift to the vicinity of 
the deconfinement transition line.
\end{abstract}

\maketitle


\section{Introduction}

The lattice QCD (LQCD) simulations have become feasible for thermal systems 
at zero quark chemical potential ($\mu$)~\cite{Kog}. 
As for $\mu^2>0$, however, lattice QCD has the well-known sign problem, and 
the results are still far from perfection; for example, see 
Ref.~\cite{Kogut2} and references therein. 

Several approaches have been proposed to solve the sign problem. 
One of them is the use of imaginary chemical potential, 
since the fermionic determinant appearing in the Euclidean partition 
function is real in the case; for example, see 
Refs.~\cite{FP,Elia,Chen,Wu,Lomb} and references therein.  
If the physical quantity such as chiral condensate is known 
in the imaginary $\mu$ region, one can extrapolate it 
to the real $\mu$ region, until there appears a discontinuity. 
Furthermore, in principle, one can evaluate with the Fourier transformation 
the canonical partition function with fixed quark number 
from the grand canonical partition function with imaginary 
$\mu$~\cite{Kratochvila,Forcrand,Alecandru}. 

Roberge and Weiss (RW)~\cite{RW} found that the partition function of 
SU($N$) gauge theory with imaginary chemical potential $\mu =i\theta/\beta$ for fermion number 
\begin{align}
Z(\theta )& = \int D\psi D\bar{\psi} DA_\mu 
\exp 
\Big[ 
- \int d^{4}x
\nonumber \\
&
\big\{ 
\bar{\psi}(\gamma D-m_0)\psi 
-{\frac{1}{4}}F_{\mu\nu}^2 
-i{\theta\over{\beta}}\bar{\psi}\gamma_4\psi
\big\}
\Big] ,
\label{eq:EQ1}
\end{align}
is a periodic function of $\theta$ with a period $2\pi/N$, that is
$Z(\theta+2{\pi}k/N)=Z(\theta)$ for any integer $k$, 
by showing that $Z(\theta+2{\pi}k/N)$ is reduced to $Z(\theta)$ with 
the ${\mathbb Z}_{N}$ transformation 
\bea
\psi \to U\psi , \quad 
A_{\nu} \to UA_{\nu}U^{-1} - {i\over{g}}(\partial_{\nu}U)U^{-1} \;,
\label{z3}
\eea
where $U(x,\tau)$ are elements of SU($N$) with 
$
U(x,\beta)=\exp(-2i \pi k/N)U(x,0).  
$
Here $\psi$ is the fermion field, $F_{\mu\nu}$ is the strength tensor of the 
gauge field $A_\nu$, and $\beta$ is the inverse of temperature $T$. 
The RW periodicity means that 
$Z(\theta)$ is invariant under the 
extended ${\mathbb Z}_{N}$ transformation 
\bea
\theta \to \theta + \tfrac{2 \pi k}{N}, \
\psi \to U\psi, \
A_{\nu} \to UA_{\nu}U^{-1} - \tfrac{i}{g} (\partial_{\nu}U)U^{-1}. 
\label{extended-z3}
\eea
Quantities invariant under the extended ${\mathbb Z}_N$ 
transformation, such as the effective potential $\Omega(\theta)$ and 
the chiral condensate, keep the RW periodicity. 
Meanwhile, the Polyakov loop $\Phi$ is not invariant under 
the transformation (\ref{extended-z3}), since 
it is transformed as 
$\Phi \to \Phi e^{-i{2\pi k/N}}$. 
In general, non-invariant quantities 
such as $\Phi$  do not have the periodicity. 
This problem can be solved by introducing the modified 
Polyakov loop $\Psi(\theta) \equiv \Phi\exp(i\theta)$ invariant under 
(\ref{extended-z3}), as shown later.

Roberge and Weiss also showed with perturbation that in the high $T$ region 
$d\Omega(\theta)/d\theta$ and $\Phi(\theta)$ are discontinuous at 
$\theta={(2k+1)\pi/N}$, and also found 
with the strong coupled lattice theory that 
the discontinuity disappears in the low $T$ region. 
This is called the Polyakov-loop RW phase transition in this paper and it 
is observed in the lattice simulations~\cite{FP,Elia,Chen,Wu,Lomb}. 

\begin{figure}[htbp]
\begin{center}
 \includegraphics[width=0.4\textwidth]{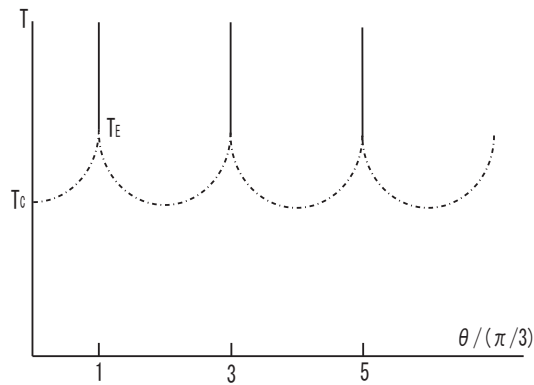} 
\end{center}
\caption{
The RW prediction on the QCD phase diagram in 
the $\theta$-$T$ plane. 
The solid lines represent the Polyakov-loop RW phase 
transitions, 
and the dot-dashed ones correspond to 
the chiral phase transitions. 
}
\label{fig1}
\end{figure}

Figure \ref{fig1} shows a predicted phase diagram in the $\theta$-$T$ plane. 
The solid lines represent the Polyakov-loop RW phase transitions, 
and the dot-dashed lines do 
the chiral phase transitions predicted by the 
lattice simulation, although the result of the simulation 
is not conclusive 
since the current quark mass taken is much heavier than 
the realistic one $5 \sim 10$~MeV. 

The Polyakov-loop RW transition  is not an ordinary 
first-order deconfinement transition. 
Although both transitions are defined by discontinuities of $\Phi$, 
the latter is a jump of the absolute value $|\Phi|$ 
from almost zero to a finite value, but the former is a discontinuity of $\Phi$ in its phase, as shown later. 
The discontinuity means that the Polyakov-loop RW transition is surely 
first-order in the phase.

As shown later, the chiral condensate $\sigma$ is also 
not smooth on the Polyakov-loop RW transition lines 
(the solid lines of Fig. \ref{fig1}) 
unless $\sigma$ is zero. 
This is called the chiral RW transition in this paper. 
This is also not an ordinary second-order chiral transition. 
The former is a jump of $d\sigma/d\theta$ from a finite value to 
its minus sign, while the latter is a divergence of $d\sigma/d\theta$.

As an approach complementary to first-principle 
lattice simulations, one can consider several effective models. 
One of them is the Nambu--Jona-Lasinio (NJL) model~\cite{NJ1}.
Although the NJL model is a useful method for understanding the chiral symmetry breaking, this model does not possess a confinement mechanism.
As a reliable model that can treat both the chiral 
and the deconfinement phase transition,  
we can consider the Polyakov loop extended NJL (PNJL) model
~\cite{Meisinger,Dumitru,Fukushima,Ghos,Megias,Ratti1,Ciminale,Ratti2,
Rossner,Hansen,Sasaki1,Schaefer,Kashiwa3,Fu} in which 
the deconfinement phase transition is described by the Polyakov loop.

The PNJL model has already been applied to the real chemical potential region 
and many interesting results, in particular on the relation between  
the chiral and the deconfinement phase transition, are 
reported~\cite{Meisinger,Dumitru,Fukushima,Ghos,Megias,
Ratti1,Ciminale,Ratti2,Rossner,Hansen,Sasaki1,Schaefer,Kashiwa3,Fu}. 
For example, the confinement mechanism shifts the critical 
endpoint of the chiral phase transition~\cite{AY,KKKN,Kashiwa} to higher $T$ 
and lower $\mu$~\cite{Rossner,Kashiwa3}. 
In order to confirm that these analyses are reliable, 
we should test the validity of the PNJL model by 
comparing the model results with the lattice ones. This is possible 
in the imaginary chemical potential region 
where the lattice simulation is feasible. 
If the PNJL model is successful in reproducing 
lattice results in the imaginary chemical potential region, 
this will imply not only that the PNJL model is reliable 
for both the real and imaginary chemical potential regions 
but also that 
the lattice results can be reasonably extrapolated 
to the real chemical potential region by using the PNJL model.

In the previous paper \cite{Sakai1}, we first applied the PNJL model to 
the imaginary chemical potential region and investigated 
the phase diagram in the chiral limit. 
Among many effective models, the PNJL model has both the chiral symmetry and 
the extended ${\mathbb Z}_3$ symmetry needed to reproduce the RW periodicity. 
We showed with the extended ${\mathbb Z}_3$ symmetry
that the Polyakov-loop RW transition is first-order in the phase of 
$\Psi$ and the chiral RW transition is second-order. 
We also showed that both the crossover deconfinement and the second-order 
chiral transition take place in the $\theta$-$T$ plane, as expected. 
In this paper, we make more extensive analyses on the four phase transitions 
by newly adding quark mass term and scalar-type 
eight-quark interaction to the PNJL Lagrangian and seeing their effects on 
the transitions. 
Throughout these analyses, we will find that 
results of the PNJL model are consistent 
with all the lattice results.

This paper is organized as follows. 
In {\S}2, we describe the PNJL model. 
In {\S}3, the extended ${\mathbb Z}_3$ symmetry, which plays a crucial role in our analyses, are introduced. 
In {\S}4, numerical results are presented. 
Section 5 is devoted to summary. 

\section{The PNJL model}

The model we consider is 
the following two-flavor PNJL Lagrangian: 
\begin{align}
 {\cal L}  =& {\bar q}(i \gamma_\nu D^\nu -m_0)q \notag\\
             &\hspace{3mm} + G_{\rm s}[({\bar q}q)^2 
                          +({\bar q}i\gamma_5 {\vec \tau}q)^2] 
              - {\cal U}(\Phi [A],{\Phi} [A]^*,T) ,
             \label{eq:E1}
\end{align}
where $q$ denotes the two-flavor quark field, 
$m_0$ does the current quark mass, and 
$D^\nu=\partial^\nu-iA^\nu-i\mu\delta^{\nu}_{0}$. 
The field $A^\nu$ is defined as 
$A^\nu=\delta^{\nu}_{0}gA^0_a{\lambda^a\over{2}}$
with the gauge field $A^\nu_a$, 
the Gell-Mann matrix $\lambda_a$ and the gauge coupling $g$.
In the NJL sector, 
${\vec \tau}$ stands for the isospin matrix, and  
$G_{\rm s}$ denotes the coupling constant of the scalar-type 
four-quark interaction. 
In {\S}4, we will add a scalar-type eight-quark \cite{Kashiwa3,Kashiwa} 
interaction to the PNJL 
Lagrangian to discuss the effect on the phase diagram. 
Effects of eight-quark interactions are discussed also in
the three flavor NJL model~\cite{Osipov1}.
The Polyakov potential ${\cal U}$, defined in (\ref{eq:E13}), 
is a function of the Polyakov loop $\Phi$ and its complex 
conjugate $\Phi^*$,
\begin{align}
\Phi      = {1\over{N_{\rm c}}}{\rm Tr} L,~~~~
\Phi^{*}  = {1\over{N_{\rm c}}} {\rm Tr}L^\dag ,
\end{align}
with
\begin{align}
L({\bf x}) = {\cal P} \exp\Bigl[
                {i\int^\beta_0 d \tau A_4({\bf x},\tau)}\Bigr],
\end{align}
where ${\cal P}$ is the path ordering and $A_4 = iA_0 $. 
In the chiral limit ($m_0=0$), 
the Lagrangian density has the exact 
$SU(2)_{\rm L} \times SU(2)_{\rm R}
\times U(1)_{\rm v} \times SU(3)_{\rm c}$  symmetry. 

The temporal component of the gauge field is diagonal 
in the flavor space, because the color and the flavor space 
are completely separated out in the present case. 
In the Polyakov gauge, $L$ can be written in a diagonal form 
in the color space~\cite{Fukushima}: 
\begin{align}
L 
=  e^{i \beta (\phi_3 \lambda_3 + \phi_8 \lambda_8)}
= {\rm diag} (e^{i \beta \phi_a},e^{i \beta \phi_b},
e^{i \beta \phi_c} ),
\label{eq:E6}
\end{align}
where $\phi_a=\phi_3+\phi_8/\sqrt{3}$, $\phi_b=-\phi_3+\phi_8/\sqrt{3}$
and $\phi_c=-(\phi_a+\phi_b)=-2\phi_8/\sqrt{3}$. 
The Polyakov loop $\Phi$ is an exact order parameter of the spontaneous 
${\mathbb Z}_3$ symmetry breaking in the pure gauge theory.
Although the ${\mathbb Z}_3$ symmetry is not an exact one 
in the system with dynamical quarks, it still seems to be a good indicator of 
the deconfinement phase transition. 
Therefore, we use $\Phi$ to define the deconfinement phase transition.

Under the mean field approximation (MFA), the Lagrangian density becomes
\begin{align}
{\cal L}_{\rm MFA} = 
{\bar q}( i \gamma_\mu D^\mu - (m_0+\Sigma_{\rm s}) )q - U_{\rm M}(\sigma) - {\cal U}(\Phi,\Phi^{*},T), 
\label{eq:E7} 
\end{align}
where
\begin{align}
\sigma = \langle \bar{q}q \rangle, ~~~~~
\Sigma_{\rm s} = -2 G_{\rm s} \sigma , ~~~~~
U_{\rm M}= G_{\rm s} \sigma^2. 
\label{eq:E10}
\end{align}

Using the usual techniques, 
one can obtain the thermodynamic potential per volume
\begin{align}
\Omega = & -{T\ln Z\over{V}}\notag\\
       = &  -2 N_f \int \frac{d^3{\rm p}}{(2\pi)^3}
          \Bigl[ 3 E ({\rm p}) + \frac{1}{\beta}
         {\rm Tr_c}\ln(1+Le^{-\beta E^-({\rm p})})  \notag\\
        &  + \frac{1}{\beta}{\rm Tr_c}\ln(1+L^\dag e^{-\beta E^+({\rm p})})
         \Bigr] +U_{\rm M}(\sigma )+{\cal U}(\Phi,\Phi^{*},T), 
\label{eq:E11}
\end{align}
where $E({\rm p})=\sqrt{{\bf p}^2+M^2}$, 
$E^\pm({\rm p})=E({\rm p})\pm \mu =E({\rm p})\pm i\theta/\beta$ and $M=m_0 + \Sigma_{\rm s}$. 
In this paper, the thermodynamic potential per volume, $\Omega$, is simply 
called the thermodynamic potential. 
After some algebra, the thermodynamic potential $\Omega$ becomes~\cite{Ratti1}
\begin{align}
\Omega =& -2 N_f \int \frac{d^3{\rm p}}{(2\pi)^3}
         \Bigl[ 3 E ({\rm p}) \nonumber\\
        & + \frac{1}{\beta}
           \ln~ [1 + 3(\Phi+\Phi^{*} e^{-\beta E^-({\bf p})}) 
           e^{-\beta E^-({\bf p})}+ e^{-3\beta E^-({\bf p})}]
         \nonumber\\
        & + \frac{1}{\beta} 
           \ln~ [1 + 3(\Phi^{*}+{\Phi e^{-\beta E^+({\bf p})}}) 
              e^{-\beta E^+({\bf p})}+ e^{-3\beta E^+({\bf p})}]
	      \Bigl]\nonumber\\
        & +U_{\rm M}+{\cal U}. 
\label{eq:E12} 
\end{align}

We use ${\cal U}$ of Ref.~\cite{Ratti1} that is fitted to a lattice QCD 
simulation 
in the pure gauge theory at finite $T$~\cite{Boyd,Kaczmarek}: 
\begin{align}
&{{\cal U}\over{T^4}} =  -\frac{b_2(T)}{2} {\Phi}^*\Phi
              -\frac{b_3}{6}({\Phi^*}^3+ \Phi^3)
              +\frac{b_4}{4}({\Phi}^*\Phi)^2, \label{eq:E13} \\
&b_2(T)   = a_0 + a_1\Bigl(\frac{T_0}{T}\Bigr)
                 + a_2\Bigl(\frac{T_0}{T}\Bigr)^2
                 + a_3\Bigl(\frac{T_0}{T}\Bigr)^3,  \label{eq:E14}
\end{align}
where parameters are summarized in Table I.  
The Polyakov potential yields a deconfinement phase transition at 
$T=T_0$ in the pure gauge theory.
In the previous paper \cite{Sakai1}, hence, 
$T_0$ was taken to be $270$ MeV predicted 
by the pure gauge lattice QCD calculation. 
However, the PNJL model with this value of $T_0$ yields somewhat larger value of the transition temperature at zero density than that predicted 
by the full LQCD simulation~\cite{Karsch3,Karsch4,MCheng}. 
Therefore, we rescale $T_0$ in {\S}4. 

\begin{table}[h]
\begin{center}
\begin{tabular}{llllll}
\hline
~~~~~$a_0$~~~~~&~~~~~$a_1$~~~~~&~~~~~$a_2$~~~~~&~~~~~$a_3$~~~~~&~~~~~$b_3$~~~~~&~~~~~$b_4$~~~~~
\\
\hline
~~~~6.75 &~~~~-1.95 &~~~~2.625 &~~~~-7.44 &~~~~0.75 &~~~~7.5 
\\
\hline
\end{tabular}
\caption{
Summary of the parameter set in the Polyakov sector
used in Ref.~\cite{Ratti1}. 
All parameters are dimensionless. 
}
\end{center}
\end{table}

The variables of $\Phi$, ${\Phi}^*$ and $\sigma$ 
satisfy the stationary conditions, 
\bea
\partial \Omega/\partial \Phi=0, \quad
\partial \Omega/\partial \Phi^{*}=0, \quad
\partial \Omega/\partial \sigma=0 . 
\label{eq:SC}
\label{condition}
\eea
The solutions of the stationary conditions do not give 
a global minimum $\Omega$ 
necessarily; there is a possibility that they yield a local minimum or even 
a maximum. We then search a global minimum directly by varying the variables 
and check the solutions to satisfy (\ref{eq:SC}). 
The physical thermodynamic potential $\Omega(\theta)$ at each $\theta$ is 
obtained by inserting the solutions at each $\theta$
into (\ref{eq:E12}).


\section{Extended ${\mathbb Z}_3$ symmetry}

The thermodynamic potential $\Omega$ of Eq. (\ref{eq:E12}) is not 
invariant under the ${\mathbb Z}_3$ transformation, 
\bea
\Phi(\theta) \to \Phi(\theta) e^{-i{2\pi k/{3}}} \;,\quad
\Phi(\theta)^{*} \to \Phi(\theta)^{*} e^{i{2\pi k/{3}}} \;, 
\eea
although 
${\cal U}$ of (\ref{eq:E13}) is invariant. 
Instead of the ${\mathbb Z}_3$ symmetry, however, 
$\Omega$ is invariant under the extended ${\mathbb Z}_3$ transformation, 
\begin{align}
&e^{\pm i \theta} \to e^{\pm i \theta} e^{\pm i{2\pi k\over{3}}},\quad  
\Phi(\theta)  \to \Phi(\theta) e^{-i{2\pi k\over{3}}}, 
\notag\\
&\Phi(\theta)^{*} \to \Phi(\theta)^{*} e^{i{2\pi k\over{3}}} .
\label{eq:K2}
\end{align}
It is convenient to introduce the modified Polyakov loop 
$\Psi \equiv e^{i\theta}\Phi$ and 
$\Psi^{*} \equiv e^{-i\theta}\Phi^{*}$ 
invariant under the transformation (\ref{eq:K2}). 
The extended ${\mathbb Z}_3$ transformation is then 
rewritten into 
\begin{align}
&e^{\pm i \theta} \to e^{\pm i \theta} e^{\pm i{2\pi k\over{3}}}, \quad
\Psi(\theta) \to \Psi(\theta), \notag\\ 
&\Psi(\theta)^{*} \to \Psi(\theta)^{*} ,
\label{eq:K2'}
\end{align}
and $\Omega$ is also into  
\begin{align}
\Omega = & -2 N_f \int \frac{d^3{\rm p}}{(2\pi)^3}
          \Bigl[ 3 E ({\rm p}) 
          + \frac{1}{\beta}\ln~ [1 + 3\Psi e^{-\beta E({\bf p})}
\notag\\
          &+ 3\Psi^{*}e^{-2\beta E({\bf p})}e^{\beta \mu_{\rm B}}
          + e^{-3\beta E({\bf p})}e^{\beta \mu_{\rm B}}]
\notag\\
          &+ \frac{1}{\beta} 
           \ln~ [1 + 3\Psi^{*} e^{-\beta E({\bf p})}
          + 3\Psi e^{-2\beta E({\bf p})}e^{-\beta\mu_{\rm B}}
\notag\\
          &+ e^{-3\beta E({\bf p})}e^{-\beta\mu_{\rm B}}]
	      \Bigl]+U_{\rm M}+\Bigl[-{b_2(T)T^4\over{2}}\Psi^{*} \Psi
\notag\\
          &-{b_3 T^4\over{6}}({\Psi^{*}}^3 e^{\beta \mu_{\rm B}}
          +\Psi^3 e^{-\beta\mu_{\rm B}})
          +{b_4T^4\over{4}}(\Psi^{*} \Psi)^2\Bigl] ,
\label{eq:K3} 
\end{align}
where $\mu_{\rm B}=3 \mu= i 3 \theta/\beta$ 
is the baryonic chemical potential and 
the factor $e^{\pm\beta\mu_{\rm B}}$ is invariant 
under the transformation (\ref{eq:K2'}). 
Obviously, $\Omega$ is invariant under the transformation 
(\ref{eq:K2'}). 

As a feature of the extended ${\mathbb Z}_3$ transformation (\ref{eq:K2'}), 
the external parameter $\mu=i\theta/\beta$ varies with (\ref{eq:K2'}), and  
the modified Polyakov-loop is invariant under (\ref{eq:K2'}) 
and then 
not an order parameter of the extended ${\mathbb Z}_3$ symmetry. 
The same is seen for the chiral transformation in 
the chiral perturbation theory~\cite{Weinberg,Gasser}. 
The Lagrangian of the theory is 
\begin{align}
L_{\rm chPT} = &{1\over{16}}F^2{\rm Tr}\left\{\partial_\mu U_{\rm
NG}\partial^\mu U_{\rm NG}^\dagger\right\}
\notag\\
  & +{1\over{2}}v{\rm Tr}\left\{M_{\rm q}(U_{\rm NG}^\dagger +U_{\rm NG})\right\}+\cdots 
\label{eq:EchPT1}
\end{align}
with 
\bea
U_{\rm NG}&=&\exp{\left(2i\sum_{a=1}^8\xi_a\lambda^a\right)}, 
\eea
where $F^2$ is a positive parameter,  and $-v$, $\xi_a$ and $M_{\rm q}$ represent, the expectation value of the quark bilinear, the Nambu-Goldstone boson fields and the real matrix for quark masses in three flavor space, respectively. 
The Lagrangian (\ref{eq:EchPT1}) is not invariant 
under the chiral transformation 
\begin{align}
U_{\rm NG}\to \exp{\left(i\sum_{a=1}^8\theta_a^{\rm R}\lambda_a\right)}
U_{\rm NG}\exp{\left(-i\sum_{a=1}^8\theta_a^{\rm L}\lambda_a\right)}, 
\label{eq:EchPT2}
\end{align}
since the second term including $M_{\rm q}$ in 
the Lagrangian breaks the symmetry explicitly, 
where $\theta_a^{\rm L}$ and $\theta_a^{\rm R}$ are arbitrary real parameters 
of the transformation for the left- and right-handed quark fields. However, 
once the quark mass matrix $M_{\rm q}$ is assumed to be transformed as 
\begin{align}
M_{\rm q} \to \exp{\left(i\sum_{a=1}^8\theta_a^{\rm R}\lambda_a\right)}
M_{\rm q}\exp{\left(-i\sum_{a=1}^8\theta_a^{\rm L}\lambda_a\right)}, 
\label{eq:EchPT3}
\end{align}
the Lagrangian (\ref{eq:EchPT1}) becomes 
invariant under the chiral transformation, and 
$v$ is not an order parameter of the extended chiral symmetry anymore.

Although we have no order parameter for the extended ${\mathbb Z}_3$ symmetry, 
this symmetry leads many useful conclusions \cite{Sakai1} as shown below. 
Under the transformation $\theta \to \theta + 2\pi k/3$, 
(\ref{eq:K3}) keeps the same form, 
if $\Psi(\theta)$ and $\Psi(\theta)^{*}$ are 
replaced by 
$\Psi(\theta +{2\pi k/3})$ and 
$\Psi(\theta +{2\pi k/3})^{*}$, respectively. 
This means that the stationary conditions for $\Psi(\theta)$ 
and $\Psi(\theta)^{*}$ 
agree with those for $\Psi(\theta +{2\pi k/3})$ and 
$\Psi(\theta +{2\pi k/3})^{*}$, respectively, and then that 
\begin{align}
\Psi(\theta +\frac{2\pi k}{3})=\Psi(\theta)
\quad {\rm and}
\quad
\Psi(\theta +\frac{2\pi k}{3})^{*}=\Psi(\theta)^{*} . 
\label{psi-RW}
\end{align}

The potential $\Omega$ depends on $\theta$ through 
$\Psi(\theta)$, $\Psi(\theta)^{*}$, $\sigma(\theta)$ 
and $e^{3i\theta}$. We then denote $\Omega(\theta)$ by  
$\Omega(\theta)=
\Omega(\Psi(\theta),\Psi(\theta)^{*},e^{3i\theta})$, 
where $\sigma(\theta)$ is suppressed since it is irrelevant to 
discussion shown below. 
The RW periodicity of $\Omega$ is then shown as 
\begin{align}
\Omega (\theta +\frac{2\pi k}{3})=&\Omega (\Psi(\theta+\frac{2\pi k}{3}),\Psi(\theta+\frac{2\pi k}{3})^{*},
e^{3i(\theta+\frac{2\pi k}{3})})\notag\\=&\Omega(\theta), 
\label{eq:K4b}
\end{align}
by using (\ref{psi-RW}) in the 
second equality. 

Equation (\ref{eq:K3}) keeps the same form 
under the transformation $\theta \to -\theta$, 
if $\Psi(\theta)$ and $\Psi(\theta)^{*}$ are replaced by 
$\Psi(-\theta)^{*}$ and $\Psi(-\theta)$, respectively. 
This indicates that 
\bea
\Psi(-\theta)=\Psi(\theta)^{*} \quad {\rm and} \quad 
\Psi(-\theta)^{*}=\Psi(\theta) .
\label{psi-z2}
\eea
Furthermore, $\Omega$ is a real function, as shown in (\ref{eq:K3}). 
Using these properties, one can show that 
\begin{align}
\Omega(\theta)=&(\Omega(\theta))^*=\Omega(\Psi(\theta)^{*},\Psi(\theta),e^{-3i\theta} )\notag\\
=&\Omega(\Psi(-\theta),\Psi(-\theta)^{*}, 
e^{-3i\theta})
=\Omega(-\theta) .
\end{align}
Thus, $\Omega$ is a periodic even function of $\theta$ with 
a period $2\pi/3$. 
The chiral condensate $\sigma(\theta)$ is also a 
periodic even function of $\theta$, 
$\sigma(\theta)=
\sigma(\theta+2 \pi k/3)=\sigma(-\theta)$, 
because it is given by $\sigma(\theta)=d\Omega(\theta)/dm_0$. 
Furthermore, the quark number density $\rho_{\rm v}=-d\Omega/d(iT\theta)$ is 
pure imaginary and a periodic odd function of $\theta$.

The modified Polyakov loop $\Psi$ has a periodicity of (\ref{psi-RW}). 
The real (imaginary) part of $\Psi$ is even (odd) under the interchange 
$\theta \leftrightarrow - \theta$, because  
\bea
{\rm Re}[\Psi(\theta)]&=&(\Psi(\theta) + \Psi(\theta)^*)/2=
{\rm Re}[\Psi(-\theta)] ,\nonumber\\
{\rm Im}[\Psi(\theta)]&=&(\Psi(\theta) - \Psi(\theta)^*)/(2i)
=-{\rm Im}[\Psi(-\theta)] \nonumber\;,
\eea
where use has been made of (\ref{psi-z2}). 
Thus, the real (imaginary) part of $\Psi$ is a periodic even (odd) function of 
$\theta$.
Similarly, the absolute value $|\Psi |$ (phase $\phi$) of the Polyakov loop is a periodic even (odd) function of $\theta$, because 
$|\Psi|=\sqrt{({\rm Re }[\Psi] )^2+({\rm Im}[\Psi] )^2}$ 
($\phi =\arctan{({{\rm Im}[\Psi]/{{\rm Re}[\Psi]}})}$). 

Since $\Omega(\theta)$, $\Psi(\theta)$ and $\sigma(\theta)$ 
are periodic functions of $\theta$ with a period $2\pi/3$, here we think a period $0 \le \theta \le 2\pi/3$. 
In the region, periodic even functions such as 
$\Omega(\theta)$, $\sigma(\theta)$, ${\rm Re}[\Psi(\theta)]$ and $|\Psi |$ are 
symmetric with respect to a line $\theta=\pi/3$. 
This indicates that such an even function has 
a cusp at $\theta=\pi/3$, 
if the gradient $d\Omega/d\theta$ is not zero. 
Such a cusp comes out in the high $T$ region, 
as shown in \S 4 with numerical calculations. 
This means that the chiral RW phase 
transition at $\theta=\pi/3$ is second order. 

Meanwhile, ${\rm Im}[\Psi(\theta)]$, $\phi$ and $\rho_{\rm v}$ 
are periodic odd functions of $\theta$. 
This leads to the fact that these are discontinuous at $\theta=\pi/3$, 
if the odd functions are not zero there. 
Thus, the Polyakov-loop RW phase transition appears as first order in 
${\rm Im}[\Psi(\theta)]$ and $\phi$ and as second order in 
${\rm Re}[\Psi(\theta)]$ and $|\Psi |$. 
The RW transition appearing in $\rho_{\rm v}$  
at $\theta=\pi/3$ is also first order. 
These are seen in the high $T$ region, as shown in \S 4. 
The orders of the Polyakov-loop and chiral RW phase transitions and 
the RW transition appearing in $\rho_{\rm v}$ 
are not affected by the existence of the current quark mass and 
multi-quark interactions introduced later, since 
the periodicity and the odd/even property of the physical quantities 
are not changed.

The dynamical variables $\Psi$ and $\Psi^*$ are 
also invariant under the continuous phase transformation, 
\begin{align}
e^{\pm i\theta} \to e^{\pm i\theta} e^{\pm i\alpha}, ~~~~~
\Phi\to \Phi e^{-i\alpha},~~~~~\Phi^*\to \Phi^*e^{i\alpha},
\label{eq:K5}
\end{align}
for an arbitrary real parameter $\alpha$. 
However, the factor $e^{\pm\beta\mu_{\rm B}}(=e^{\pm3i\theta})$ 
and the potential $\Omega$ 
including the factor are not invariant. 
If $\Omega$ were invariant under (\ref{eq:K5}), 
the continuous symmetry would lead to a simple relation 
$\Psi (\theta +\alpha )=\Psi (\theta )$, that is 
$\Phi ( \theta + \alpha )=e^{-i\alpha}\Phi (\theta)$, 
that guarantees that $\Phi$ is a smooth function of $\theta$. 
When $T$ is small under the condition that $\mu$ is imaginary and 
$\Psi$ and $\Psi^*$ are not zero, 
the thermodynamic potential (\ref{eq:K3}) is reduced to 
\begin{align}
\Omega\sim & 
 -2 N_f \int \frac{d^3{\rm p}}{(2\pi)^3}
         \Bigl[ 3 E ({\rm p}) + \frac{1}{\beta} \ln~ [1 + 3\Psi e^{-\beta E({\bf p})}]
\notag\\ +& \frac{1}{\beta} \ln~ [1 + 3\Psi^{*} e^{-\beta E({\bf p})}]
	      \Bigl] +U_{\rm M}-{a_3{T_0}^3T\over{2}}\Psi^*\Psi  .
\label{eq:K3a} 
\end{align}
This has no explicit $\mu_{\rm B}(=3i\theta T)$ dependence. 
Therefore, at low temperature, $\Omega$ is 
approximately invariant under (\ref{eq:K5}) and 
$\Phi$ can rotate smoothly as $\theta$ varies. 
At high temperature, however, effects of the explicit $\mu_{\rm B}$ dependence are not negligible 
and $\Phi$ can not rotate smoothly. 
Thus, it is obvious that the Polyakov-loop RW phase transition at high $T$ 
is originated in the factor $e^{\pm\beta\mu_{\rm B}}$ in (\ref{eq:K3}). 
At high temperature, the continuous symmetry 
under the transformation (\ref{eq:K5}) is broken into a discrete symmetry, 
i.e., the extended ${\mathbb Z}_3$ symmetry, 
through the factor $e^{\pm\beta\mu_{\rm B}}$. 

It is easily seen that the ordinary NJL model respects 
the chiral symmetry but it does not preserve 
the extended ${\mathbb Z}_3$ symmetry. 
On the contrary, the 3-dimensional 3-state Potts model~\cite{Patel,DeGrand,Alford,Kim} respects the extended ${\mathbb Z}_3$ symmetry 
and then has the RW periodicity, 
but it does not possess the chiral symmetry, 
since the model is a paradigm of QCD in the large quark mass limit. 
In lattice QCD (LQCD), to avoid the quadratic divergence, the chemical potential should be introduced just like the fourth component of an imaginary constant vector potential~\cite{Hasenfratz}, i.e., $e^{\mu a}U_4$ or $e^{-\mu a}U_4^\dag$, where $a$ and $U_4(=e^{iaA_4})$ are the lattice spacing and the fourth component of gauge field on lattice, respectively. 
In this case, the RW periodicity is expected to be naturally satisfied. 
In Table II, we summarize these symmetry properties in 
the three effective models together with QCD and LQCD. 
Among the effective models, only the PNJL model 
has the same properties as QCD does. 

\begin{table}[h]
\begin{center}
\begin{tabular}{ccccc}
\hline
 Theory& Chiral& ${\mathbb Z}_3$& extended ${\mathbb
 Z}_3 $ & RW periodicity 
\\
\hline
 QCD& P & PP & P & P
\\
\hline
 LQCD & P & PP & P & P 
\\
\hline
 NJL & P & ND & B & B
\\
\hline
 PNJL & P & PP & P & P
\\
\hline
 3-d 3-state Potts~~~~~& B & PP & P & P 
\\
\hline
\end{tabular}
\caption{
Summary of symmetry properties in several theories. 
In QCD, LQCD, NJL and PNJL, we consider the chiral limit, $m_0=0$; 
P means ``preserved", PP does ``preserved in the pure gauge sector", 
B does ``broken explicitly" and ND does ``not defined". 
}
\end{center}
\end{table}

Although the original NJL model does not has the extended ${\mathbb Z}_3$ 
symmetry, it may be a good approximation to the PNJL model. 
In the mean field approximation, the fermionic part $\Omega_{\rm NJL}^{\rm f}$ of the NJL thermodynamic potential is given by 
\begin{align}
\Omega_{\rm NJL}^{\rm f}=&-2 N_fN_c  \int \frac{d^3{\rm p}}{(2\pi)^3}
         \Bigl[ E ({\rm p}) 
\notag\\
        & + \frac{1}{\beta}\ln~ [1 + e^{-\beta E^-({\bf p})}]
         + \frac{1}{\beta}\ln~ [1 + e^{-\beta E^+({\bf p})}]
         \Bigl]. \quad
\label{eq:K7} 
\end{align}
It is easily seen that 
$\Omega_{\rm PNJL}^{\rm f}(\Phi =\Phi^*=1)=\Omega_{\rm NJL}^{\rm f}$. 
Therefore, the fermionic part of the NJL thermodynamic potential coincides with that of the PNJL model in the perfectly deconfinement phase. 
In the confinement phase, meanwhile, 
$\Omega_{\rm PNJL}^{\rm f}(\Phi =\Phi^*=0)=\Omega_{\rm NJL}^{\rm f}$ 
is not held in general. 
Actually, it is also easily shown that 
\begin{align}
&\Omega_{\rm PNJL}^{\rm f}(\Phi =\Phi^*=0) =
\notag\\& {\Omega_{\rm NJL}^{\rm f} (T,\mu )
+\Omega_{\rm NJL}^{\rm f} (T,\mu +{2\pi i \over{3}}T)
+\Omega_{\rm NJL}^{\rm f} (T,\mu -{2\pi i \over{3}}T)\over{N}}. 
\label{eq:K9}
\end{align}
Therefore, in the confined phase, the fermionic part of the PNJL thermodynamic potential does not coincide with $\Omega_{\rm NJL}^{\rm f}$ but 
with the ${\mathbb Z}_3$ symmetrized average of $\Omega_{\rm NJL}^{\rm f}$. 
However, in the zero temperature limit, the thermodynamic potential of the PNJL model is reduced to that of the NJL model as 
\begin{align}
\Omega&_{\rm PNJL} (T=0)
\notag\\&= -6N_f \int \frac{d^3{\rm p}}{(2\pi)^3}
           \Bigl[E ({\bf p}) -\theta (-E^- ({\bf p}))E^- ({\bf p}) \Bigl]
\notag\\
           &~~+U_{\rm M}(\sigma ,\rho_{\rm v}(T\to 0,\mu,\sigma ))+{\cal
	   U}(T\to 0,\Phi,\Phi^*)
\notag\\&=\Omega_{\rm NJL}(T=0). 
\label{eq:E12-2}
\end{align}
Therefore, the difference between the NJL model and the PNJL model 
is significant only in the intermediate temperature region. 
In both the high and the low temperature region, 
the two model give similar results 
for the physics concerning the chiral symmetry. 
It should be also remarked that the vacuum term in the PNJL thermodynamic potential (\ref{eq:E12}) is the same as that of the NJL model. 
Therefore, the parameters included in the NJL sector of the PNJL model 
are the same as the parameters in the original NJL model, if they are determined phenomenologically at $T=\mu =0$. 
This also ensures that the original NJL model is a good approximation 
to the PNJL model. 
Thus, the difference between the two models is significant 
in the intermediate temperature region. Actually, the PNJL model makes 
the critical endpoint~\cite{AY,KKKN,Kashiwa} shift 
to higher $T$ and lower $\mu$ 
than the NJL model does~\cite{Rossner,Kashiwa3}.


\section{Numerical Results}

Since the NJL model is nonrenormalizable, it is then needed to 
introduce a cutoff in the momentum integration. 
Here we take the three-dimensional momentum cutoff 
\begin{equation}
\int \frac{d^3{\bf p}}{(2 \pi)^3}\to 
{1\over{2\pi^2}} \int_0^\Lambda dp p^2.
\label{eq:E15}
\end{equation}
Hence, the present model has three parameters 
$m_0$, $\Lambda$, $G_{\rm s}$ in the NJL sector. 
Following Ref.~\cite{Kashiwa}, we use $\Lambda =0.6315$ GeV and $G_{\rm s}=5.498$ GeV$^{-2}$, but vary $m_0$ as a free parameter. 
If we put $m_0=5.5$MeV, these parameters reproduce 
the pion decay constant $f_{\pi}=93.3$MeV and the pion mass $M_{\pi}=138$MeV.  

\begin{figure}[htbp]
\begin{center}
 \includegraphics[width=0.4\textwidth]{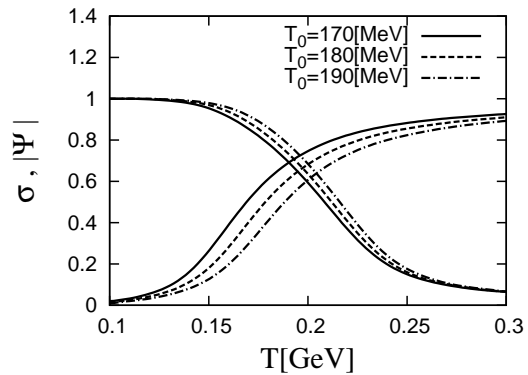} 
\end{center}
\caption{$T$ dependence of the chiral condensate normalized by 
$\sigma (T=0,\mu =0)$ and 
the absolute value of the modified Polyakov loop $\Psi(\theta)$ 
at $\theta =0$ for three cases of $T_0$. 
Increasing (decreasing) functions of $\theta$ denote 
$|\Psi|$ ($\sigma$) for all the cases. 
}
\label{fig2}
\end{figure}

Figure \ref{fig2} shows the $T$ dependence of the chiral condensate $\sigma$ 
and the absolute value of the modified Polyakov loop, $|\Psi|$, 
at $\theta =0$ for three cases of $T_0$. 
The quantity $|\Psi|$ ($\sigma$) indicates 
that a crossover transition takes place at $T_{\rm D}=240$MeV ($T_{\rm C}=261$MeV) 
for $T_0=270$MeV and at $T_{\rm D}=176$MeV ($T_{\rm C}=221$MeV) 
for $T_0=190$MeV. 
We took the original value $T_0=270$MeV 
in the previous work~\cite{Sakai1}, but in this paper 
we take the rescaled one $T_0=190$MeV. 
In this case the values of $T_{\rm C}$ and $T_{\rm D}$  
are closer to 170$\sim$180MeV, the results of 
the two-flavor full LQCD simulation~\cite{Karsch3,Karsch4,MCheng}, than 
those in the case of $T_0=270$MeV, 
although the difference $T_{\rm C}-T_{\rm D}$ becomes larger 
as $T_0$ decreases, as shown in Fig. \ref{fig2}. 

\begin{figure}[htbp]
\begin{center}
 \includegraphics[width=0.4\textwidth]{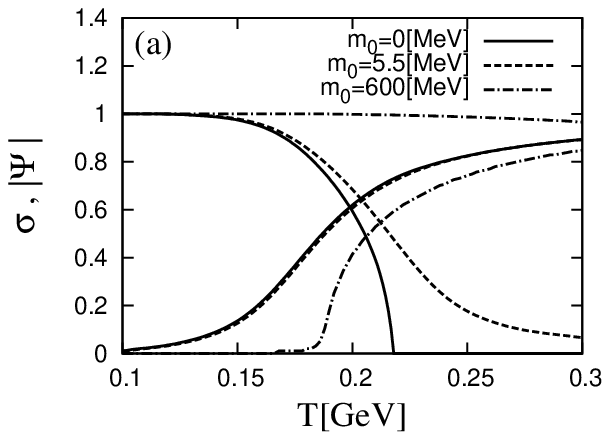} 
 \includegraphics[width=0.4\textwidth]{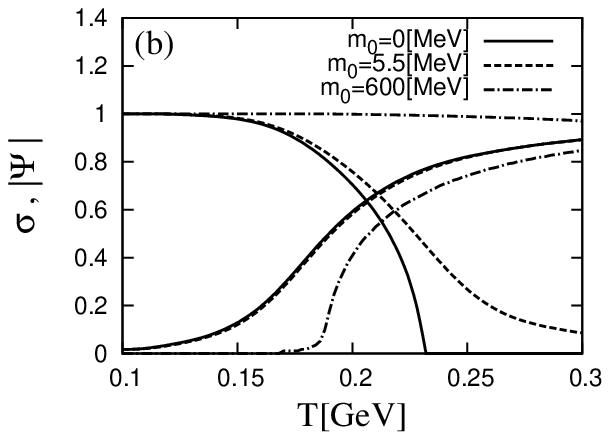} 
\end{center}
\caption{The chiral condensate $\sigma$ normalized by $\sigma (T=0,\mu=0)$ 
and the absolute value of $\Psi(\theta)$ 
as a function of $T$ for three cases of $m_0$; 
(a) corresponds to $\theta =0$ and (b) to $\theta ={\pi\over{6}}$.
Increasing (decreasing) functions of $T$ denote 
$|\Psi|$ ($\sigma$) for all the cases. 
 }
\label{fig3}
\end{figure}

Figure \ref{fig3} shows the $T$ dependence of the chiral condensate and the modified Polyakov loop for three cases of  $m_0$. 
In the chiral limit $m_0=0$, the chiral condensate shows 
that a second-order chiral phase transition takes place around 
230~MeV not only for $\theta =0$ but also for $\theta \neq 0$.  
The transition becomes crossover when $m_0$ is finite. 
Meanwhile, it is found from the modified Polyakov loop that 
the crossover deconfinement transition that appears around 180~MeV 
becomes sharper as $m_0$ increases. 
This behavior is consistent with the result of the 3-dimensional 3-state Potts model~\cite{DeGrand,Patel,Alford,Kim} in which
a first-order deconfinement transition is observed for 
large quark mass. 

\begin{figure}[htbp]
\begin{center}
 \includegraphics[width=0.4\textwidth]{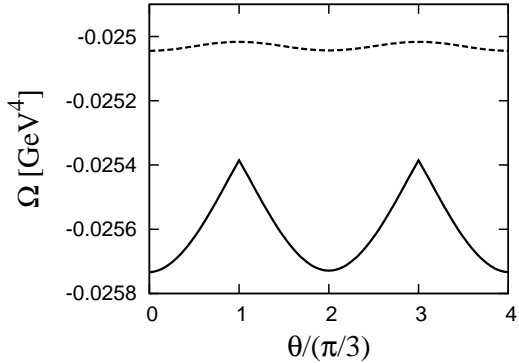} 
\end{center}
\caption{Thermodynamic potential $\Omega$ (in GeV$^4$) 
as a function of $\theta$. 
The dashed line corresponds to the case of $T=170$~MeV and 
the solid one to that of $T=200$~MeV. 
}
\label{fig4}
\end{figure}

Figure \ref{fig4} shows the thermodynamic potential $\Omega$ as a function of $\theta$ in two cases of $T=170$ and 200~MeV. 
The potential $\Omega$ is smooth everywhere 
in the low $T$ case, but not at $\theta =(2k+1)\pi/{3}$ 
in the high $T$ case. 
This result is consistent with 
the RW prediction \cite{RW} and lattice simulation~\cite{Lomb}
on the $\theta$ and the $T$ dependence 
of the QCD thermodynamic potential. 
Qualitative features of $\Omega$ at $\theta =(2k +1)\pi /3$ are the same 
as in the chiral limit case~\cite{Sakai1}. 

\begin{figure}[htbp]
\begin{center}
 \includegraphics[width=0.4\textwidth]{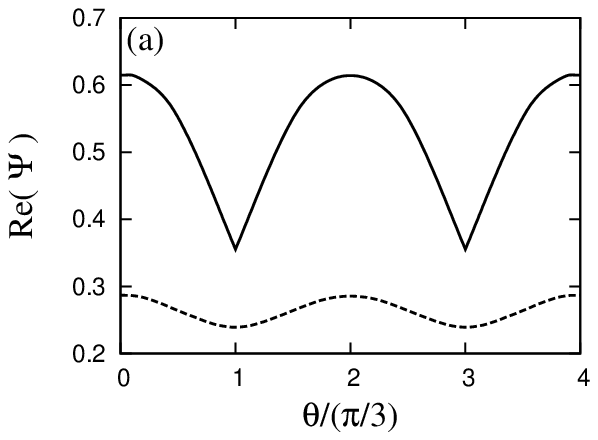} 
 \includegraphics[width=0.4\textwidth]{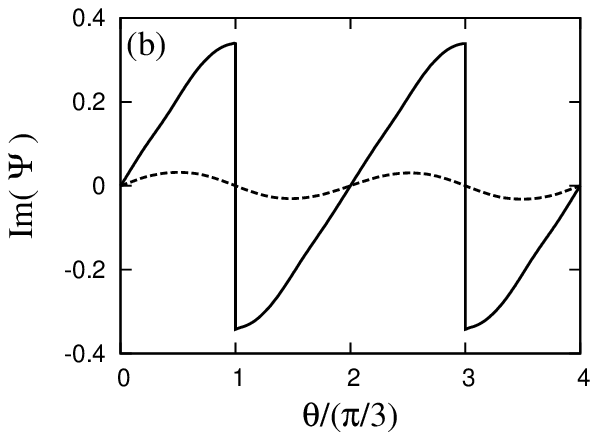} 
\end{center}
\caption{The modified Polyakov loop $\Psi(\theta)$ 
as a function of $\theta$; 
(a) for the real part and (b) for the imaginary part. 
Definitions of lines are the  same as in Fig.~4. }
\label{fig5}
\end{figure}

Figure \ref{fig5} shows the real and imaginary parts of 
$\Psi(\theta)$ as a function of $\theta$. 
In the case of $T=200$~MeV, the imaginary part of 
$\Psi(\theta)$ is discontinuous at $\theta =(2k+1)\pi/{3}$, 
while the real part of $\Psi(\theta)$ is continuous but 
not smooth there. 
The transition appearing at $\theta =(2k+1)\pi/{3}$ 
is the Polyakov loop RW phase transition. 
In the case of $T=170$~MeV, meanwhile, both the real and the imaginary part 
are smooth everywhere. 

\begin{figure}[htbp]
\begin{center}
 \includegraphics[width=0.4\textwidth]{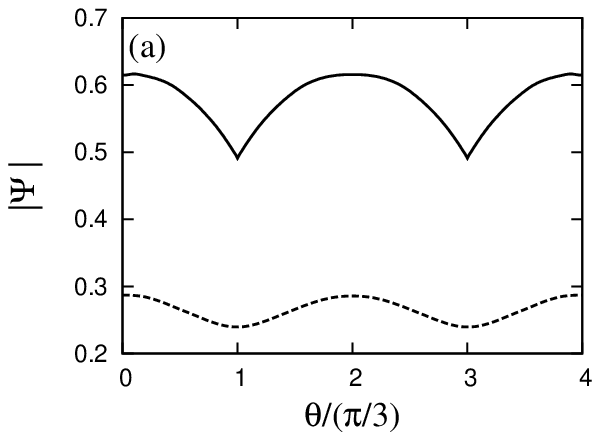} 
 \includegraphics[width=0.4\textwidth]{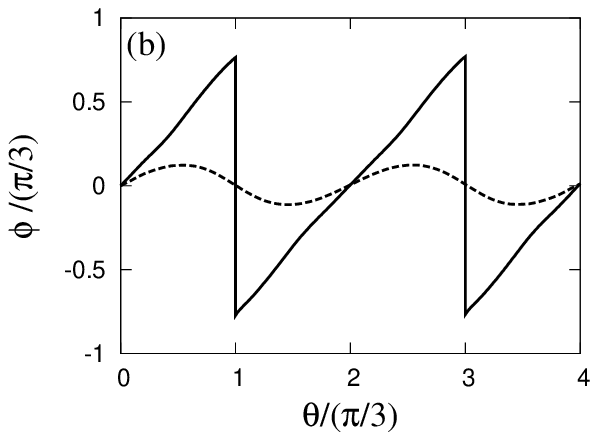} 
\end{center}
\caption{The modified Polyakov loop $\Psi(\theta)$ 
as a function of $\theta$; 
(a) for the absolute value and (b) for the phase. 
Definitions of lines are the  same as in Fig.~4. }
\label{fig6}
\end{figure}

Figure \ref{fig6} shows the absolute value $|\Psi |$ and the phase $\phi$ of 
the modified Polyakov loop as a function of $\theta$. 
In the case of $T=200$~MeV, the phase $\phi (\theta)$ is discontinuous at $\theta =(2k+1)\pi/{3}$, while the absolute value $|\Psi (\theta)|$ is continuous but not smooth there. 
In the case of $T=170$~MeV, meanwhile, both the absolute value and the phase are smooth everywhere. 
All the results on the $\theta$ and the $T$ dependence of $\Psi$ are 
consistent with the results of lattice simulations~\cite{FP,Elia,Chen,Wu,Lomb}. 
As an interesting feature, 
the transition appears as first order 
in the phase $\phi$ and the imaginary part of $\Psi(\theta)$
and as second order in the absolute value and the real part of $\Psi(\theta)$. 
Qualitative features of the Polyakov-loop RW transition are 
the same as in the chiral limit~\cite{Sakai1}. 

\begin{figure}[htbp]
\begin{center}
 \includegraphics[width=0.4\textwidth]{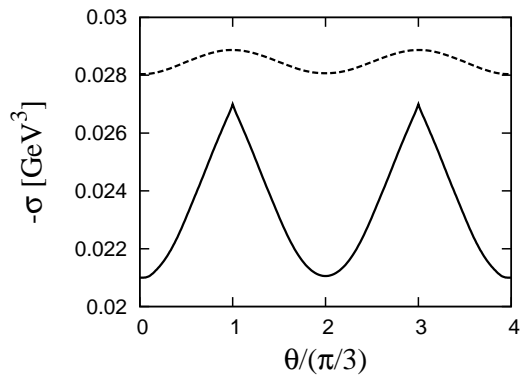} 
\end{center}
\caption{Chiral condensate $\sigma$ (in GeV$^3$) as a function of $\theta$.
Definitions of lines are the  same as in Fig.~4. }
\label{fig7}
\end{figure}

Figure \ref{fig7} 
shows the chiral condensate $\sigma$ as a function of $\theta$. 
In the case of $T=200$~MeV, $\sigma$ has a cusp at each of lines 
$\theta =(2k+1)\pi/{3}$. Thus, the chiral phase transition of second order 
comes out at $\theta =(2k+1)\pi/{3}$. 
Meanwhile, in the case of $T=170$MeV, 
there is no cusp at $\theta =(2k+1)\pi/{3}$, indicating no 
chiral phase transition there. 
As a feature not seen in the chiral limit, 
$\sigma$ and the cusp do not vanish even in much higher temperature 
than $T=200$~MeV. 

\begin{figure}[htbp]
\begin{center}
 \includegraphics[width=0.4\textwidth]{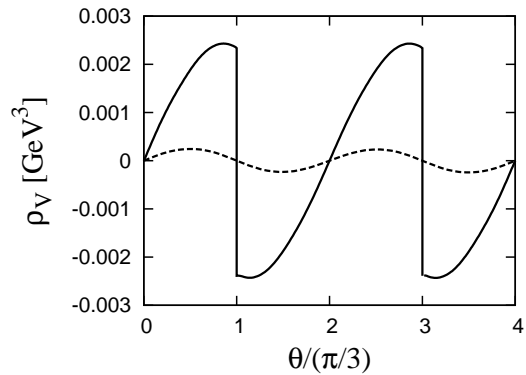} 
\end{center}
\caption{Imaginary part of the quark number density $\rho_{\rm v}$ (in GeV$^3$) as a function of $\theta$.
Definitions of lines are the  same as in Fig.~4. }
\label{fig8}
\end{figure}

Figure \ref{fig8} shows the imaginary part of 
quark number density $\rho_{\rm v}$ as a function $\theta$;
note that the real part is always zero. 
In the case of $T=200$~MeV, it is discontinuous at $\theta =(2k+1)\pi/{3}$, 
indicating that the phase transition is first order. 
In the case of $T=200$~MeV, $\rho_{\rm v}$ is smooth everywhere. 
The $\theta$ and the $T$ dependence of $\rho_{\rm v}$ are  
consistent with the results of lattice simulations~\cite{Elia}. 
Comparing Fig. \ref{fig8} with Fig. \ref{fig5}(b), we see that 
the imaginary part of the quark number density has $\theta$ 
dependence similar to the imaginary part of the modified Polyakov loop. 
This is natural, because 
the former is related to the fourth component of the vector current, 
while the latter to the fourth component of the vector field.  

Thus, the PNJL results are consistent 
with the lattice ones \cite{Elia,Chen,Wu}, 
except that the temperature difference 
$T_{\rm C}-T_{\rm D}$ is considerably large in the former 
but small in the latter. 
Since the lattice simulations made in the imaginary $\mu$ region 
have small lattice sizes, the results are not conclusive necessarily. 
However, it is important to check whether the PNJL model can 
reproduce the present lattice data in the finite imaginary $\mu$ region. 
This is discussed below from a qualitative point of view.

Our calculations have no free parameter, since  
the parameters of the Polyakov-potential sector have been fixed 
to reproduce the results of the lattice QCD simulations and 
the other parameters in the NJL sector have been adjusted to reproduce 
the empirical values of $f_\pi$ and $M_\pi$, i.e., 
$f_\pi =93.3$MeV and $M_\pi =138$MeV. 
Therefore, we need a new parameter to improve this situation. 
It is reported that the value of the critical temperature is sensitive 
to the strength of scalar-type eight-quark interaction 
in both the NJL model~\cite{Kashiwa} and the PNJL one~\cite{Kashiwa3}. 
Furthermore, the NJL and PNJL models with the eight-quark interaction 
can reproduce lighter $\sigma$-meson mass that may be 
phenomenologically more favorable~\cite{Kashiwa3,Kashiwa}. 
Therefore, we can consider the scalar-type eight-quark 
interaction as a reasonable extension of the ordinary PNJL model.

\begin{figure}[htbp]
\begin{center}
 \includegraphics[width=0.4\textwidth]{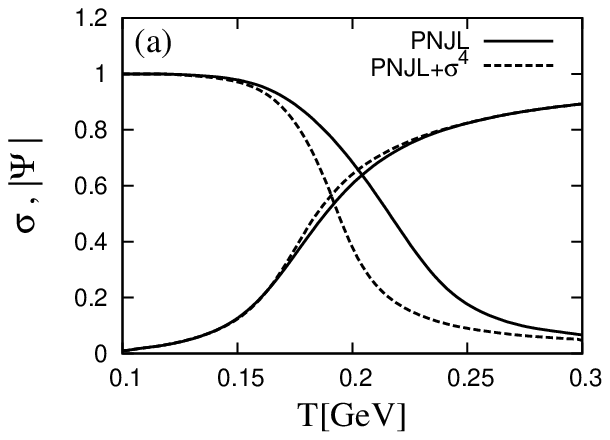} 
 \includegraphics[width=0.4\textwidth]{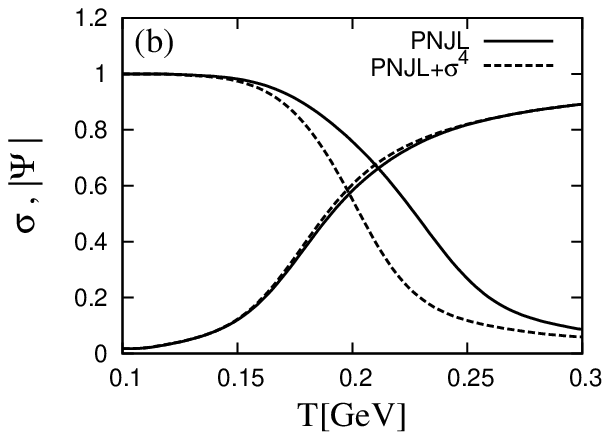} 
\end{center}
\caption{The chiral condensate $\sigma$ normalized by $\sigma (T=0,\mu =0)$ 
and the absolute value of $\Psi(\theta)$ as a function of $T$ 
with and without the eight-quark interaction; 
(a) for $\theta =0$ and (b) for $\theta ={\pi\over{6}}$. 
Increasing (decreasing) functions of $\theta$ denote 
$|\Psi|$ ($\sigma$) for all the cases. 
}
\label{fig9}
\end{figure}

We add the eight-quark interaction~\cite{Kashiwa3,Kashiwa} 
\begin{align}
G_{\rm s8}[(\bar{q}q)^2+(\bar{q}i\gamma_5\vec{\tau}q)^2]^2
\label{eq:E1a}
\end{align}
to Lagrangian (\ref{eq:E1}). 
The scalar part $\Sigma_{\rm s}$ of the quark self-energy and 
the mesonic potential $U_{\rm M}$ are modified into 
\begin{align}
\Sigma_{\rm s}=-2G_{\rm s}\sigma -4G_{\rm s8}\sigma^3,~~~~~
U_{\rm M}=G_{\rm s}\sigma^2+3G_{\rm s8}\sigma^4. 
\label{eq:E9a}
\end{align}
Figure \ref{fig9} shows the $T$ dependence of the chiral condensate and the modified Polyakov loop with and without the eight-quark interaction. 
We set the parameter $G_{\rm s8}$ to 270.9~GeV$^{-8}$ to 
reproduce $m_\sigma =600$~MeV at $T=\theta =0$~\cite{Kashiwa}, 
where $m_\sigma$ is the $\sigma$-meson mass. 
We see for $\theta =0$ that the eight-quark interaction makes 
the chiral transition temperature $T_{\rm C}$ shift down to 
the deconfinement transition temperature $T_{\rm D}\sim 180$~MeV. 
Similar effect is seen also for $\theta ={\pi\over{6}}$. 

\begin{figure}[htbp]
\begin{center}
 \includegraphics[width=0.4\textwidth]{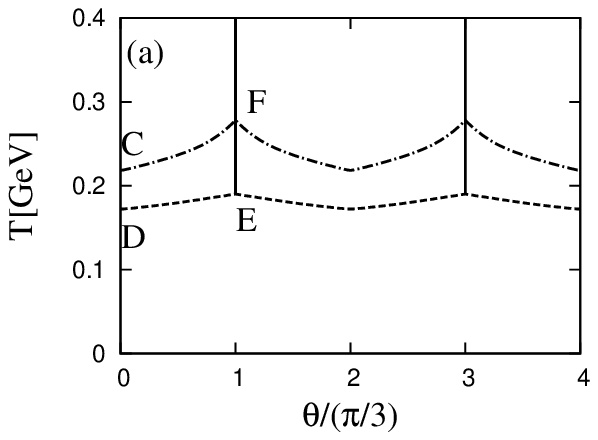}
 \includegraphics[width=0.4\textwidth]{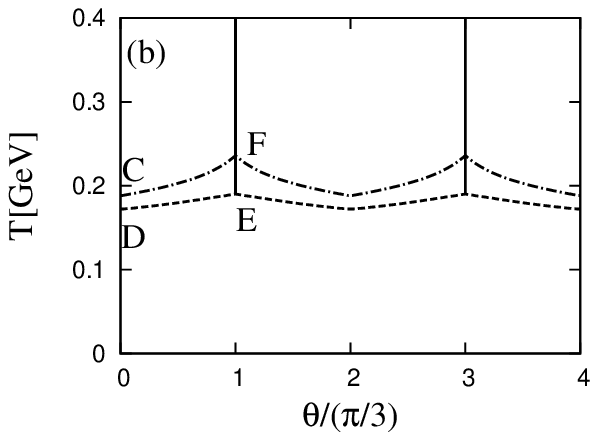}
\end{center}
\caption{
The phase diagram in the $\theta$-$T$ plane in the case of $m_0=0$; 
the eight-quark interaction is switched off in (a) and on in (b). 
The solid vertical line starting from point 
E represents the Polyakov-loop RW phase transition. 
The dashed curve between D and E represents 
the deconfinement transition of crossover, and 
the dot-dashed curve between C and F does the second-order 
chiral phase transition. 
The second-order chiral RW transition occurs on the line between 
E and F. 
}
\label{fig10}
\end{figure}

Figure \ref{fig10} represents the phase diagram in the $\theta$-$T$ plane 
in the case of $m_0=0$. 
The phase diagram is symmetric with respect to 
each of lines $\theta=k \pi/3$ for any integer $k$. 
The dashed curve between D and E represents 
the deconfinement transition of crossover, and 
the dot-dashed curve between C and F does the second-order 
chiral phase transition. 
For $\theta \neq k \pi/3$, thus, 
the chiral phase transition occurs at higher $T$ than 
the deconfinement transition does. 
The solid vertical line starting from point 
E represents the Polyakov-loop RW phase transition. 
The chiral RW transition of second order occurs on 
the line between E and F. Above F, the chiral condensate is zero. 
Point F turns out to be a bifurcation of 
the chiral phase transition line, and point E is the endpoint
of both the Polyakov-loop RW  and chiral RW transitions. 
The chiral phase transition line between C and F is shifted down to 
the vicinity of the deconfinement transition line between D and E by adding 
the eight-quark interaction. 
In the case without (with) the eight-quark interaction, 
the temperatures at the point C, D, E, F are $T_{\rm C}=221(188)$~MeV, 
$T_{\rm D}=172(172)$~MeV, $T_{\rm E}=190(190)$~MeV 
and $T_{\rm F}=280(236)$~MeV, 
respectively.

\begin{figure}[htbp]
\begin{center}
 \includegraphics[width=0.4\textwidth]{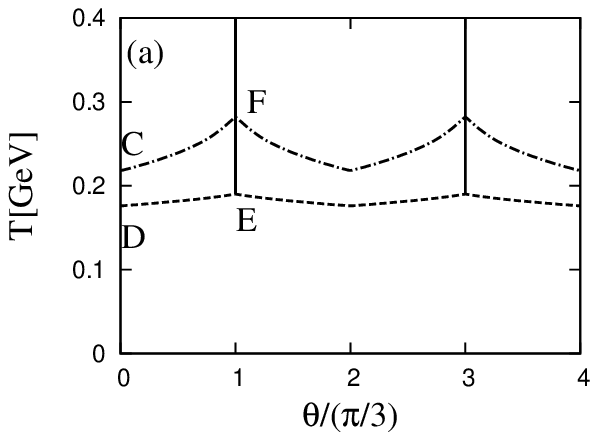} 
 \includegraphics[width=0.4\textwidth]{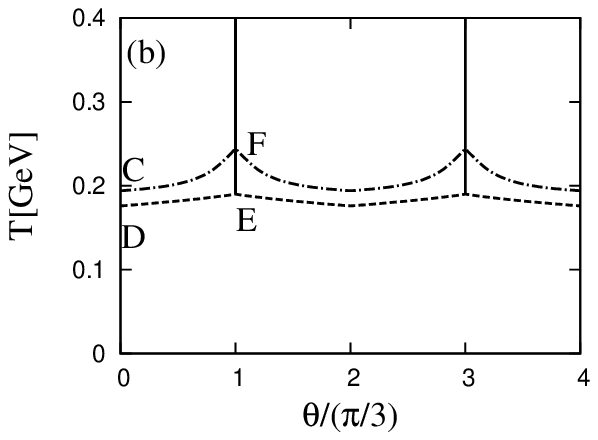} 
\end{center}
\caption{
The phase diagram in the $\theta$-$T$ plane in the case of $m_0=5.5$~MeV; 
the eight-quark interaction is switched off in (a) and on in (b). 
The solid vertical line starting from point 
E represents both the Polyakov-loop RW  phase transition  
and the chiral RW phase transition of second order. 
The dashed curve between D and E represents 
the deconfinement transition of crossover and 
the dot-dashed curve between C and F does the chiral transition of crossover. 
 }
\label{fig11}
\end{figure}

Figure \ref{fig11} represents the phase diagram in the $\theta$-$T$ plane 
in the case of $m_0=5.5$~MeV. 
Again, the phase diagram is symmetric with respect to 
each of lines $\theta=k \pi/3$. 
The dashed curve between D and E represents 
the deconfinement transition of crossover, and 
the dot-dashed curve between C and F does the chiral transition of crossover. 
Also for this finite $m_0$ case, 
the chiral transition occurs at higher $T$ than 
the deconfinement transition. 
The solid vertical line starting from point 
E represents both the Polyakov-loop RW phase transition and 
the chiral RW phase transition of second order. 
Point E is the endpoint of both the RW transitions. 
As a viewpoint different from the $m_0=0$ case~\cite{Sakai1}, 
point F is neither a bifurcation of the chiral phase transition line 
nor an endpoint of the chiral RW phase transition, 
since the chiral condensate $\sigma$ is always finite on the 
vertical line $\theta=\pi/3$, that is, also on the line above point F. 
Just as in the case of $m_0=0$, 
the chiral transition line between C and F is shifted down to the vicinity of 
the deconfinement transition line between D and E by adding 
the eight-quark interaction. 
In the case without (with) the eight-quark interaction, 
the temperatures at the point C, D, E, F are $T_{\rm C}=218(194)$~MeV, 
$T_{\rm D}=176(176)$~MeV, $T_{\rm E}=190(190)$~MeV 
and $T_{\rm F}=282(244)$~MeV, respectively.

\begin{figure}[htbp]
\begin{center}
 \includegraphics[width=0.29\textwidth,angle=-90]{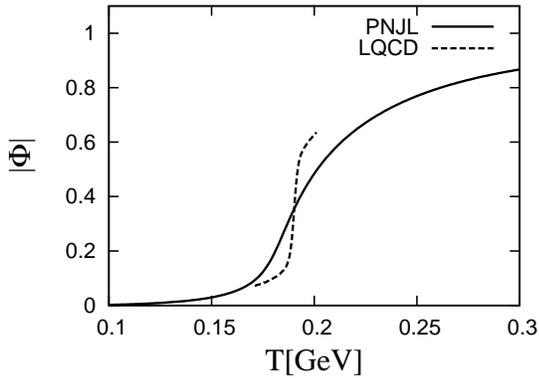} 
\end{center}
\caption{
The absolute value of $\Phi$ as a function of $T$ plane in the case 
of $\theta/(\pi/3)=0.96$. 
The solid curve is the result of PNJL with $m_0=100$~MeV, while 
the dashed one is the result  of lattice QCD \cite{Wu}. 
 }
\label{fig12}
\end{figure}

Finally, we make a brief trial of the PNJL extrapolation, that is,  
the lattice data obtained in the imaginary $\mu$ region 
is extrapolated with PNJL to the real $\mu$ region. 
The absolute value of the Polaykov loop, $|\Phi|$, is measured for 
the two-flavor case in Ref. \cite{Wu}, 
although the lattice simulation of Wilson dynamical quarks 
has a small size of $8^3 \times 4$. 
As the first step, we try to reproduce the lattice data \cite{Wu} with PNJL. 
The screening mass of quark is estimated to be a few hundreds MeV 
in the simulation, 
so we simply change only $m_0$ to 100 MeV in PNJL calculations done above 
with $\Lambda =0.6315$ GeV, $G_{\rm s}=5.498$ GeV$^{-2}$, $T_0=190$~MeV 
and $G_{\rm s8}=0$. 
Figure \ref{fig12} presents $|\Phi|$ as a function of $T$ in the case of 
$\theta/(\pi/3) =0.96$, where the solid (dashed) curve corresponds to 
the result of PNJL (LQCD). From a qualitative point of view, 
both the calculations give the same property that the Polykov-loop 
phase transition is crossover also for this case. 
From a quantitative point of view, however,  
LQCD gives a sharper transition than PNJL.

\begin{figure}[htbp]
\begin{center}
 \includegraphics[width=0.29\textwidth,angle=-90]{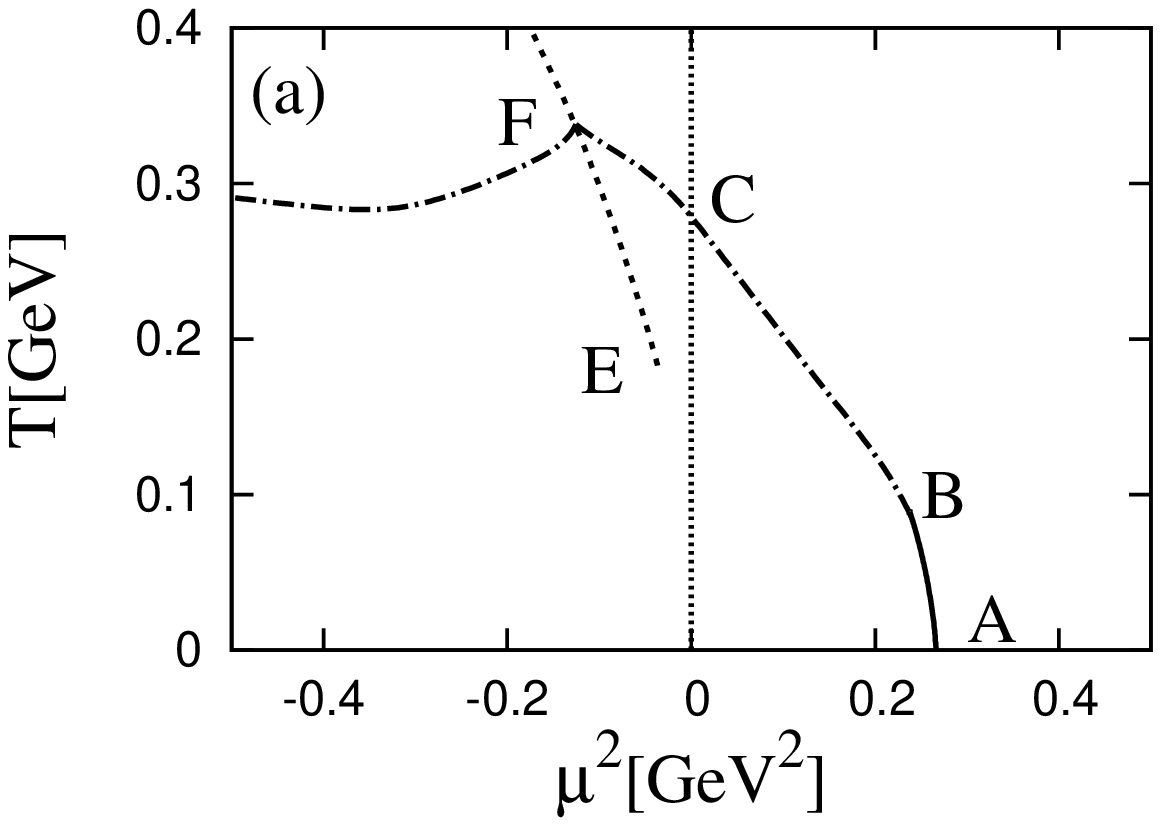} 
 \includegraphics[width=0.29\textwidth,angle=-90]{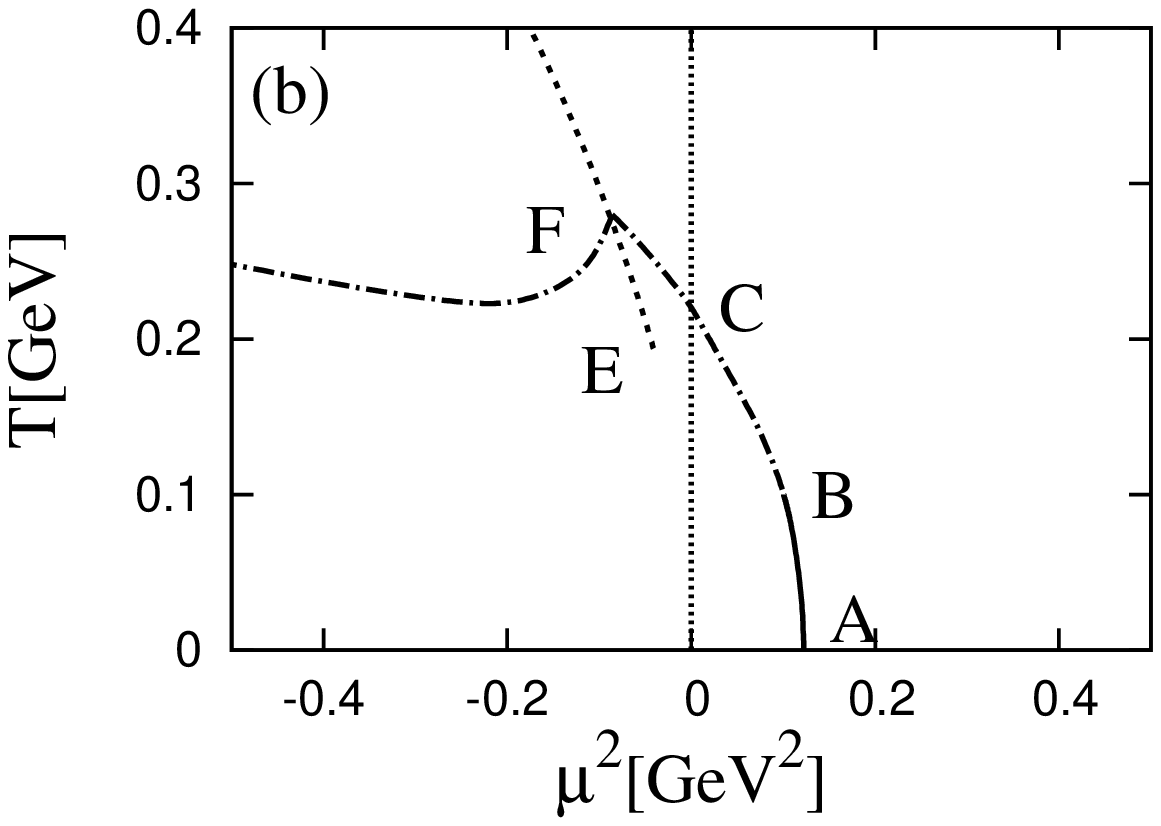}
\end{center}
\caption{
The phase diagram in the $\mu^2$-$T$ plane for 
(a) $m_0=100$~MeV and (b) $m_0=5.5$~MeV. 
The solid (dot-dashed) curves denote 
the first-order (crossover) chiral phase transitions, while the 
dashed curve does the chiral RW phase transiton of the second order. 
}
\label{fig13}
\end{figure}

Although more careful and extensive reproduction is necessary in future, now 
we do the PNJL calculation in both the real and imaginary $\mu$ 
regions with the parameter set determined above. 
Figure \ref{fig13}(a) presents the chiral phase transition curves 
in the $\mu^2$-$T$ plane. 
The solid (dot-dashed) curves represent that 
the phase transitions are the first-order (crossover), while the 
dashed curve does the chiral RW phase transition of 
the second-order. 
The curve from point F to point A is smooth, as expected. 
For comparison, in Fig. \ref{fig13}(b) we present 
the result of $m_0=5.5$~MeV, 
where the other parameters are fixed.  
Thus, the phase diagram is largely affected by the change of $m_0$. 
This result encourages us to do more accurate LQCD similations 
with larger lattice sizes and small current quark mass and 
also to make more extensive comparison 
of PNJL calculations with LQCD ones. 
In the forthcoming paper, we will make such a PNJL extrapolation 
for the case of four flavor, 
since most LQCD simulations \cite{Elia} are done in the case.

\begin{figure}[htbp]
\begin{center}
 \includegraphics[width=0.29\textwidth,angle=-90]{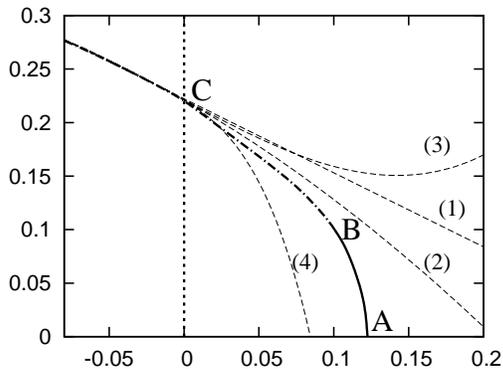}
\end{center}
\caption{
The phase diagram in the $\mu^2$-$T$ plane based on 
the extrapolation of (\ref{extra}). 
Dashed curves labeled (1)-(4)  
correspond to four cases of $m=1,2,3,4$ in (\ref{extra}), 
respectively.
}
\label{fig14}
\end{figure}

In the lattice simulations \cite{Elia,Chen,Wu}, the extrapolaton 
is made in a simple way in which 
the function 
\beq
T=\sum_{n=0}^{m} a_n \mu^{2n} 
\label{extra}
\eeq
is assumed for the chiral transition curve and 
the coefficients $a_n$ are determined so as to reproduce the lattice data 
in the imaginary $\mu$ region. 
In this paper the coefficients $a_n$ are adjusted to 
the PNJL result of Fig. \ref{fig13}(b) in the imaginary $\mu$ region. 
In Fig. \ref{fig14}, four dashed curves labeled (1)-(4) 
represent results of the simple extrapolation for $m=1,2,3,4$, respectively. 
The dashed curve of $m=4$ still deviates from the PNJL curve 
in the real $\mu$ region. Thus, the PNJL curve 
includes higher-order terms than $m=5$ that 
the simple extrapolation can not follow accurately. 
Furthermore, the simple extrapolation can not predict the position of the 
critical endpoint B. 
Thus, the PNJL extrapolaiton is superior to the simple extrapolation.

\section{Summary}

In summary, we have studied the Polyakov loop extended Nambu--Jona-Lasinio 
(PNJL) model with imaginary chemical potential. 
The phase diagram in the $\theta$-$T$ plane is studied in detail. 
Since the PNJL model possesses 
an extended ${\mathbb Z}_{3}$ symmetry, 
quantities invariant under the symmetry such as 
the thermodynamic potential, the chiral condensate and 
the modified Polyakov loop, automatically have 
the Roberge-Weiss periodicity that QCD does. 
There appear four types of phase transitions; 
Polyakov-loop RW transition, 
chiral RW transition of second order, deconfinement and chiral transitions.
The orders of the two RW transitions appearing at $\theta=(2k+1)\pi/3$ are 
determined by the RW periodicity and the even/odd property of 
the modified Polyakov loop and the chiral condensate. 
As an interesting feature, 
the Polyakov-loop RW transition comes out 
as first order in the imaginary part and the phase of the modified Polyakov 
loop, but as second order in the real part and the absolute value 
of the modified Polyakov loop. 
These results are more informative than 
the RW prediction \cite{RW} and 
the results of lattice QCD~\cite{FP,Elia,Chen,Wu,Lomb}.

We have investigated effects of current quark mass $m_0$ on the phase diagram. 
The orders of the ordinary chiral and deconfinement transitions 
depend on the presence or absence of current quark mass, as expected. 
In contrast, 
the orders of the Polyakov-loop RW and the chiral RW transition are not 
affected by the presence or absence of the current quark mass. 
The bifurcation point of the chiral transition line, which appears in the phase diagram in the chiral limit, disappears in the case of finite $m_0$. 
We have also studied effects of the scalar-type eight-quark interaction 
on the phase diagram in the $\theta$-$T$ plane. 
It is found that the eight-quark interaction makes 
the chiral transition temperature shift down to the vicinity of 
the deconfinement transition temperature in the whole range of $\theta$.

The success of the PNJL model comes from the fact 
that the PNJL model has the extended ${\mathbb Z}_{3}$ symmetry, more 
precisely that the thermodynamic potential (\ref{eq:K3}) is a function only of 
variables, $\Psi$, $\Psi^*$, $e^{\pm \beta\mu_{\rm B}}$ and $\sigma$, 
invariant under the extended ${\mathbb Z}_{3}$ symmetry. 
A reliable effective theory of QCD proposed in future 
is expected to have the same property 
in its thermodynamic potential. 
This may be a good guiding principle 
to elaborate an effective theory of QCD.

Throughout all the present analyses, 
we can confirm that the results of the present model 
are consistent with the lattice results. In the analyses of 
the chiral limit, the present model is even more informative than the lattice 
simulation. 
In this paper, we have compared our results with lattice ones 
only qualitatively, since lattice simulations 
in the imaginary chemical potential region have been done mainly for the 
four-flavor case. In the forthcoming paper, we will 
make quantitative comparison of the present model with lattice QCD. 
Determining the parameters of our model directly 
from the lattice results in the imaginary chemical potential region, 
we will extrapolate the lattice results to the real chemical potential 
region by using the present model.

\bigskip

\noindent
\begin{acknowledgments}
The authors thank M. Matsuzaki, H. Fujii, M. Tachibana, K. Miura and T. Murase for useful discussions and suggestions. 
H.K. also thanks M. Imachi and H. Yoneyama for useful discussions on the RW phase transition. 
This work has been supported in part by the Grants-in-Aid for Scientific Research (18540280) of Education, Science, Sports, and Culture of Japan.
\end{acknowledgments}



\begin{thebibliography}{19}
\expandafter\ifx\csname natexlab\endcsname\relax\def\natexlab#1{#1}\fi
\expandafter\ifx\csname bibnamefont\endcsname\relax
  \def\bibnamefont#1{#1}\fi
\expandafter\ifx\csname bibfnamefont\endcsname\relax
  \def\bibfnamefont#1{#1}\fi
\expandafter\ifx\csname citenamefont\endcsname\relax
  \def\citenamefont#1{#1}\fi
\expandafter\ifx\csname url\endcsname\relax
  \def\url#1{\texttt{#1}}\fi
\expandafter\ifx\csname urlprefix\endcsname\relax\def\urlprefix{URL }\fi
\providecommand{\bibinfo}[2]{#2}
\providecommand{\eprint}[2][]{\url{#2}}

%

\bibitem[{\citenamefont{Kogut et~al.}(1983)\citenamefont{Kogut, Stone, {H. W.
  Wyld}, {W. R. Gibbs}, Shigemitsu, {S. H. Shenker}, and {D. K.
  Sinclair}}}]{Kog}
\bibinfo{author}{\bibfnamefont{J.}~\bibnamefont{Kogut}},
  \bibinfo{author}{\bibfnamefont{M.}~\bibnamefont{Stone}},
  \bibinfo{author}{\bibnamefont{{H. W. Wyld}}},
  \bibinfo{author}{\bibnamefont{{W. R. Gibbs}}},
  \bibinfo{author}{\bibfnamefont{J.}~\bibnamefont{Shigemitsu}},
  \bibinfo{author}{\bibnamefont{{S. H. Shenker}}}, \bibnamefont{and}
  \bibinfo{author}{\bibnamefont{{D. K. Sinclair}}}, \bibinfo{journal}{Phys.\
  Rev.\ Lett.} \textbf{\bibinfo{volume}{50}}, \bibinfo{pages}{393}
  (\bibinfo{year}{1983}).

\bibitem[{\citenamefont{Kogut}(2007)}]{Kogut2}
\bibinfo{author}{\bibfnamefont{J.}~\bibnamefont{B.}~\bibnamefont{Kogut}} 
\bibnamefont{and} 
\bibinfo{author}{\bibfnamefont{D.}~\bibnamefont{K.}~\bibnamefont{Sinclair}}  
\bibinfo{howpublished}{arXiv:hep-lat/0712.2625} (\bibinfo{year}{2007}). 

\bibitem[{\citenamefont{Forcrand and Philipsen}(2002)}]{FP}
\bibinfo{author}{\bibfnamefont{P.}~\bibnamefont{de}~\bibnamefont{Forcrand}} 
\bibnamefont{and}
\bibinfo{author}{\bibfnamefont{O.}~\bibnamefont{Philipsen}},  
\bibinfo{journal}{Nucl. Phys. } \textbf{\bibinfo{volume}{B642}},
\bibinfo{pages}{290} (\bibinfo{year}{2002});
\bibinfo{author}{\bibfnamefont{P.}~\bibnamefont{de}~\bibnamefont{Forcrand}} 
\bibnamefont{and}
\bibinfo{author}{\bibfnamefont{O.}~\bibnamefont{Philipsen}},  
\bibinfo{journal}{Nucl. Phys. } \textbf{\bibinfo{volume}{B673}},
\bibinfo{pages}{170} (\bibinfo{year}{2003}). 

\bibitem[{\citenamefont{Elia and Lombardo}(2003)}]{Elia}
\bibinfo{author}{\bibfnamefont{M.}~\bibnamefont{D'Elia}} \bibnamefont{and}
\bibinfo{author}{\bibfnamefont{M.}~\bibfnamefont{P.}~\bibnamefont{Lombardo}},  
\bibinfo{journal}{Phys. Rev.\  D} \textbf{\bibinfo{volume}{67}},
\bibinfo{pages}{014505} (\bibinfo{year}{2003});
\bibinfo{author}{\bibfnamefont{M.}~\bibnamefont{D'Elia}} \bibnamefont{and}
\bibinfo{author}{\bibfnamefont{M.}~\bibfnamefont{P.}~\bibnamefont{Lombardo}},  
\bibinfo{journal}{Phys. Rev.\ D} \textbf{\bibinfo{volume}{70}},
\bibinfo{pages}{074509} (\bibinfo{year}{2004}). 

\bibitem[{\citenamefont{Chen and Luo}(2005)}]{Chen}
\bibinfo{author}{\bibfnamefont{H.}~\bibfnamefont{S.}~\bibnamefont{Chen}}
\bibnamefont{and}
\bibinfo{author}{\bibfnamefont{X.}~\bibfnamefont{Q.}~\bibnamefont{Luo}},  
\bibinfo{journal}{Phys. Rev.} \textbf{\bibinfo{volume}{D72}},
\bibinfo{pages}{034504} (\bibinfo{year}{2005})

\bibitem[{\citenamefont{Chen and Luo}(2005)}]{Wu}
\bibinfo{author}{\bibfnamefont{L.}~\bibfnamefont{K.}~\bibnamefont{Wu}}, 
\bibinfo{author}{\bibfnamefont{X.}~\bibfnamefont{Q.}~\bibnamefont{Luo}},  
\bibnamefont{and}
\bibinfo{author}{\bibfnamefont{H.}~\bibfnamefont{S.}~\bibnamefont{Chen}}, 
\bibinfo{journal}{Phys. Rev.} \textbf{\bibinfo{volume}{D76}},
\bibinfo{pages}{034505} (\bibinfo{year}{2007}).  


\bibitem[{\citenamefont{Laombardo}(2006)}]{Lomb}
\bibinfo{author}{\bibfnamefont{M.}~\bibfnamefont{P.}~\bibnamefont{Lombardo}}, 
\bibinfo{howpublished}{arXiv:hep-lat/0612017} (\bibinfo{year}{2006}). 

\bibitem[{\citenamefont{Kratochvila and Forcrand}(2004)}]{Kratochvila}
\bibinfo{author}{\bibfnamefont{S.}~\bibnamefont{Kratochvila}} \bibnamefont{and}
\bibinfo{author}{\bibfnamefont{P.}~\bibfnamefont{de}~\bibnamefont{Forcrand}},  
\bibinfo{journal}{Prog. Theor. Phys. Suppl.} \textbf{\bibinfo{volume}{153}},
\bibinfo{pages}{330} (\bibinfo{year}{2004}). 

\bibitem[{\citenamefont{Forcrand and Kratochvila}(2006)}]{Forcrand}
\bibinfo{author}{\bibfnamefont{P.}~\bibnamefont{de}~\bibnamefont{Forcrand}} \bibnamefont{and}
\bibinfo{author}{\bibfnamefont{S.}~\bibnamefont{Kratochvila}},  
\bibinfo{journal}{Nucl. Phys. {\bf B} (Proc. Suppl.) } \textbf{\bibinfo{volume}{153}},
\bibinfo{pages}{62} (\bibinfo{year}{2006}). 

\bibitem[{\citenamefont{Alexandru et al.}(2005)}]{Alecandru}
\bibinfo{author}{\bibfnamefont{A.}~\bibnamefont{Alexandru}}, 
\bibinfo{author}{\bibfnamefont{M.}~\bibnamefont{Faber}},
\bibinfo{author}{\bibfnamefont{I.}~\bibnamefont{Harva\'{a}th}}, 
\bibnamefont{and}
\bibinfo{author}{\bibfnamefont{K.}~\bibnamefont{F.}~\bibnamefont{Liu}},  
\bibinfo{journal}{Phys. Rev.} \textbf{\bibinfo{volume}{D72}},
\bibinfo{pages}{114513-1} (\bibinfo{year}{2005}). 

\bibitem[{\citenamefont{Roberge and Weiss}(1986)}]{RW}
\bibinfo{author}{\bibfnamefont{A.}~\bibnamefont{Roberge}} \bibnamefont{and}
\bibinfo{author}{\bibfnamefont{N.}~\bibnamefont{Weiss}},  
\bibinfo{journal}{Nucl. Phys. } \textbf{\bibinfo{volume}{B275}},
\bibinfo{pages}{734} (\bibinfo{year}{1986}). 

\bibitem[{\citenamefont{Nambu and Jona-Lasinio}(1961{\natexlab{a}})}]{NJ1}
\bibinfo{author}{\bibfnamefont{Y.}~\bibnamefont{Nambu}} \bibnamefont{and}
  \bibinfo{author}{\bibfnamefont{G.}~\bibnamefont{Jona-Lasinio}},
  \bibinfo{journal}{Phys.\ Rev.} \textbf{\bibinfo{volume}{122}},
  \bibinfo{pages}{345} (\bibinfo{year}{1961}); 
  \bibinfo{journal}{Phys.\ Rev.} \textbf{\bibinfo{volume}{124}},
  \bibinfo{pages}{246} (\bibinfo{year}{1961}).

\bibitem[{\citenamefont{Meisinger et al.}(1996)}]{Meisinger}
\bibinfo{author}{\bibfnamefont{P.}~\bibnamefont{N.}}~\bibnamefont{Meisinger},
\bibnamefont{and}
\bibinfo{author}{\bibfnamefont{M.}~\bibnamefont{C.}}~\bibnamefont{Ogilvie},  
  \bibinfo{journal}{Phys. Lett.\ B} \textbf{\bibinfo{volume}{379}},
  \bibinfo{pages}{163} (\bibinfo{year}{1996}). 

\bibitem[{\citenamefont{Dumitru}(2002)}]{Dumitru}
\bibinfo{author}{\bibfnamefont{A.}~\bibnamefont{Dumitru}},
\bibnamefont{and}
\bibinfo{author}{\bibfnamefont{R.}~\bibfnamefont{D.}~\bibnamefont{Pisarski}},  
\bibinfo{journal}{Phys.\ Rev.\  D} \textbf{\bibinfo{volume}{66}},
\bibinfo{pages}{096003} (\bibinfo{year}{2002}); 
\bibinfo{author}{\bibfnamefont{A.}~\bibnamefont{Dumitru}},
\bibinfo{author}{\bibfnamefont{Y.}~\bibnamefont{Hatta}},
\bibinfo{author}{\bibfnamefont{J.}~\bibnamefont{Lenaghan}},
\bibinfo{author}{\bibfnamefont{K.}~\bibnamefont{Orginos}},
\bibnamefont{and}
\bibinfo{author}{\bibfnamefont{R.}~\bibfnamefont{D.}~\bibnamefont{Pisarski}},  
\bibinfo{journal}{Phys.\ Rev.\  D} \textbf{\bibinfo{volume}{70}},
\bibinfo{pages}{034511} (\bibinfo{year}{2004}); 
\bibinfo{author}{\bibfnamefont{A.}~\bibnamefont{Dumitru}},
\bibinfo{author}{\bibfnamefont{R.}~\bibfnamefont{D.}~\bibnamefont{Pisarski}},  
\bibnamefont{and}
\bibinfo{author}{\bibfnamefont{D.}~\bibnamefont{Zschiesche}},  
\bibinfo{journal}{Phys.\ Rev.\  D} \textbf{\bibinfo{volume}{72}},
\bibinfo{pages}{065008} (\bibinfo{year}{2005}).

\bibitem[{\citenamefont{Fukushima}(2004)}]{Fukushima}
\bibinfo{author}{\bibfnamefont{K.}~\bibnamefont{Fukushima}}, 
  \bibinfo{journal}{Phys. Lett.\ B} \textbf{\bibinfo{volume}{591}},
  \bibinfo{pages}{277} (\bibinfo{year}{2004}). 

\bibitem[{\citenamefont{{S. K. Ghosh} et al.}(2006)}]{Ghos}
\bibinfo{author}{\bibnamefont{{S. K. Ghosh}}},
  \bibinfo{author}{\bibnamefont{{T. K. Mukherjee}}},
  \bibinfo{author}{\bibnamefont{{M. G. Mustafa}}}, \bibnamefont{and}
  \bibinfo{author}{\bibfnamefont{R.}~\bibnamefont{Ray}},
  \bibinfo{journal}{Phys.\ Rev.\ D} \textbf{\bibinfo{volume}{73}},
  \bibinfo{pages}{114007} (\bibinfo{year}{2006}). 

\bibitem[{\citenamefont{Megias et al.}(2006)}]{Megias}
\bibinfo{author}{\bibfnamefont{E.}~\bibnamefont{Meg{$\acute{\i}$}as}},
\bibinfo{author}{\bibfnamefont{E.}~\bibnamefont{R.}~\bibnamefont{Arriola}},
\bibnamefont{and}
\bibinfo{author}{\bibfnamefont{L.}~\bibnamefont{L.}~\bibnamefont{Salcedo}},  
  \bibinfo{journal}{Phys. Rev.\ D} \textbf{\bibinfo{volume}{74}},
  \bibinfo{pages}{065005} (\bibinfo{year}{2006}). 

\bibitem[{\citenamefont{Ratti et al.}(2006)}]{Ratti1}
\bibinfo{author}{\bibfnamefont{C.}~\bibnamefont{Ratti}},
\bibinfo{author}{\bibfnamefont{M.}~\bibfnamefont{A.}~\bibnamefont{Thaler}},
\bibnamefont{and}
\bibinfo{author}{\bibfnamefont{W.}~\bibnamefont{Weise}},  
  \bibinfo{journal}{Phys. Rev.\ D} \textbf{\bibinfo{volume}{73}},
  \bibinfo{pages}{014019} (\bibinfo{year}{2006}). 

\bibitem[{\citenamefont{Ciminale}(2007)}]{Ciminale}
\bibinfo{author}{\bibfnamefont{M.}~\bibnamefont{Ciminale}},
\bibinfo{author}{\bibfnamefont{R.}~\bibnamefont{Gatto}},
\bibinfo{author}{\bibfnamefont{N.}~\bibfnamefont{D.}~\bibnamefont{Ippolito}},
\bibinfo{author}{\bibfnamefont{G.}~\bibnamefont{Nardulli}},  
\bibnamefont{and}
\bibinfo{author}{\bibfnamefont{M.}~\bibnamefont{Ruggieri}},
  \bibinfo{journal}{Phys. Rev.\ D} \textbf{\bibinfo{volume}{77}},
  \bibinfo{pages}{054023} (\bibinfo{year}{2008});
\bibinfo{author}{\bibfnamefont{M.}~\bibnamefont{Ciminale}},
\bibinfo{author}{\bibfnamefont{G.}~\bibnamefont{Nardulli}},  
\bibinfo{author}{\bibfnamefont{M.}~\bibnamefont{Ruggieri}},
\bibnamefont{and}
\bibinfo{author}{\bibfnamefont{R.}~\bibnamefont{Gatto}},
  \bibinfo{journal}{Phys.\ Lett.\ B} \textbf{\bibinfo{volume}{657}},
  \bibinfo{pages}{64} (\bibinfo{year}{2007}).

\bibitem[{\citenamefont{Ratti et al.}(2007)}]{Ratti2}
\bibinfo{author}{\bibfnamefont{C.}~\bibnamefont{Ratti}},
\bibinfo{author}{\bibfnamefont{S.}~\bibnamefont{R\"{o}{\ss}ner}},
\bibinfo{author}{\bibfnamefont{M.}~\bibfnamefont{A.}~\bibnamefont{Thaler}},
\bibnamefont{and}
\bibinfo{author}{\bibfnamefont{W.}~\bibnamefont{Weise}},  
  \bibinfo{journal}{Eur. Phys. J.\ C} \textbf{\bibinfo{volume}{49}},
  \bibinfo{pages}{213} (\bibinfo{year}{2007}). 

\bibitem[{\citenamefont{Rossner et al.}(2007)}]{Rossner}
\bibinfo{author}{\bibfnamefont{S.}~\bibnamefont{R\"{o}{\ss}ner}},
\bibinfo{author}{\bibfnamefont{C.}~\bibnamefont{Ratti}},
\bibnamefont{and}
\bibinfo{author}{\bibfnamefont{W.}~\bibnamefont{Weise}},  
  \bibinfo{journal}{Phys. Rev.\ D} \textbf{\bibinfo{volume}{75}},
  \bibinfo{pages}{034007} (\bibinfo{year}{2007}). 

\bibitem[{\citenamefont{Hansen et al.}(2007)}]{Hansen}
\bibinfo{author}{\bibfnamefont{H.}~\bibnamefont{Hansen}}, 
\bibinfo{author}{\bibfnamefont{W.}~\bibfnamefont{M.}~\bibnamefont{Alberico}},
\bibinfo{author}{\bibfnamefont{A.}~\bibnamefont{Beraudo}}, 
\bibinfo{author}{\bibfnamefont{A.}~\bibnamefont{Molinari}},
\bibinfo{author}{\bibfnamefont{M.}~\bibnamefont{Nardi}},
\bibnamefont{and}
\bibinfo{author}{\bibfnamefont{C.}~\bibnamefont{Ratti}}, 
  \bibinfo{journal}{Phys. Rev.\ D} \textbf{\bibinfo{volume}{75}},
  \bibinfo{pages}{065004} (\bibinfo{year}{2007}). 

\bibitem[{\citenamefont{Sasaki et al.}(2007)}]{Sasaki1}
\bibinfo{author}{\bibfnamefont{C.}~\bibnamefont{Sasaki}},
\bibinfo{author}{\bibfnamefont{B.}~\bibnamefont{Friman}},
\bibnamefont{and}
\bibinfo{author}{\bibfnamefont{K.}~\bibnamefont{Redlich}}, 
\bibinfo{journal}{Phys. Rev.\ D} \textbf{\bibinfo{volume}{75}},
  \bibinfo{pages}{074013} (\bibinfo{year}{2007}). 

\bibitem[{\citenamefont{Schaefer}(2007)}]{Schaefer}
\bibinfo{author}{\bibfnamefont{B.}~\bibfnamefont{J.}~\bibnamefont{Schaefer}},
\bibinfo{author}{\bibfnamefont{J.}~\bibfnamefont{M.}~\bibnamefont{Pawlowski}},
\bibnamefont{and}
\bibinfo{author}{\bibfnamefont{J.}~\bibnamefont{Wambach}},  
  \bibinfo{journal}{Phys.\ Rev.\  D} \textbf{\bibinfo{volume}{76}},
  \bibinfo{pages}{074023} (\bibinfo{year}{2007}).

\bibitem[{\citenamefont{Kashiwa et al}(2007)}]{Kashiwa3}
\bibinfo{author}{\bibfnamefont{K.}~\bibnamefont{Kashiwa}}, 
\bibinfo{author}{\bibfnamefont{H.}~\bibnamefont{Kouno}}, 
\bibinfo{author}{\bibfnamefont{M.}~\bibnamefont{Matsuzaki}}, 
\bibnamefont{and}
\bibinfo{author}{\bibfnamefont{M.}~\bibnamefont{Yahiro}}, 
\bibinfo{journal}{Phys. Lett.\ B} \textbf{\bibinfo{volume}{662}},
\bibinfo{pages}{26} (\bibinfo{year}{2008}). 

\bibitem[{\citenamefont{Fu}(2007)}]{Fu}
\bibinfo{author}{\bibfnamefont{W.}~\bibfnamefont{J.}~\bibnamefont{Fu}},
\bibinfo{author}{\bibfnamefont{Z.}~\bibnamefont{Zhang}},
\bibnamefont{and}
\bibinfo{author}{\bibfnamefont{Y.}~\bibfnamefont{X.}~\bibnamefont{Liu}},
  \bibinfo{journal}{Phys.\ Rev.\  D} \textbf{\bibinfo{volume}{77}},
  \bibinfo{pages}{014006} (\bibinfo{year}{2008}).

\bibitem[{\citenamefont{Asakawa and Yazaki}(1989)}]{AY}
\bibinfo{author}{\bibfnamefont{M.}~\bibnamefont{Asakawa}} \bibnamefont{and}
  \bibinfo{author}{\bibfnamefont{K.}~\bibnamefont{Yazaki}},
  \bibinfo{journal}{Nucl.\ Phys.} \textbf{\bibinfo{volume}{A504}},
  \bibinfo{pages}{668} (\bibinfo{year}{1989}). 

\bibitem[{\citenamefont{Kitazawa et al. }(2002)}]{KKKN}
\bibinfo{author}{\bibfnamefont{M.}~\bibnamefont{Kitazawa}},
{\bibfnamefont{T.}~\bibnamefont{Koide}},
{\bibfnamefont{T.}~\bibnamefont{Kunihiro}},
\bibnamefont{and}
\bibinfo{author}{\bibfnamefont{Y.}~\bibnamefont{Nemoto}},
\bibinfo{journal}{Prog. Theor. Phys.} \textbf{\bibinfo{volume}{108}},
\bibinfo{pages}{929} (\bibinfo{year}{2002}). 

\bibitem[{\citenamefont{Kashiwa et al}(2006)}]{Kashiwa}
\bibinfo{author}{\bibfnamefont{K.}~\bibnamefont{Kashiwa}}, 
\bibinfo{author}{\bibfnamefont{H.}~\bibnamefont{Kouno}}, 
\bibinfo{author}{\bibfnamefont{T.}~\bibnamefont{Sakaguchi}}, 
\bibinfo{author}{\bibfnamefont{M.}~\bibnamefont{Matsuzaki}}, 
\bibnamefont{and}
\bibinfo{author}{\bibfnamefont{M.}~\bibnamefont{Yahiro}},
\bibinfo{journal}{Phys. Lett.\ B} \textbf{\bibinfo{volume}{647}},
\bibinfo{pages}{446} (\bibinfo{year}{2007}); 
\bibinfo{author}{\bibfnamefont{K.}~\bibnamefont{Kashiwa}}, 
\bibinfo{author}{\bibfnamefont{M.}~\bibnamefont{Matsuzaki}}, 
\bibinfo{author}{\bibfnamefont{H.}~\bibnamefont{Kouno}}, 
\bibnamefont{and}
\bibinfo{author}{\bibfnamefont{M.}~\bibnamefont{Yahiro}},
\bibinfo{journal}{Phys. Lett.\ B} \textbf{\bibinfo{volume}{657}},
\bibinfo{pages}{143} (\bibinfo{year}{2007}). 

\bibitem[{\citenamefont{Sakai}(2008)}]{Sakai1}
\bibinfo{author}{\bibfnamefont{Y.}~\bibnamefont{Sakai}}, 
\bibinfo{author}{\bibfnamefont{K.}~\bibnamefont{Kashiwa}}, 
\bibinfo{author}{\bibfnamefont{H.}~\bibnamefont{Kouno}}, 
\bibnamefont{and}
\bibinfo{author}{\bibfnamefont{M.}~\bibnamefont{Yahiro}},
\bibinfo{journal}{Phys. Rev. \ D} \textbf{\bibinfo{volume}{77}}
\bibinfo{page} {051901(R)}(\bibinfo{year}{2008}). 

\bibitem[{\citenamefont{Osipov et al. }(2006)}]{Osipov1}
\bibinfo{author}{\bibfnamefont{A.}~\bibfnamefont{A.}~\bibnamefont{Osipov}},
{\bibfnamefont{B.}~\bibnamefont{Hiller}},
\bibnamefont{and} 
{\bibfnamefont{J.}~\bibnamefont{da Provid\^encia}},
\bibinfo{journal}{Phys. Lett.\ B} \textbf{\bibinfo{volume}{634}},
\bibinfo{pages}{48} (\bibinfo{year}{2006});
\bibinfo{author}{\bibfnamefont{A.}~\bibfnamefont{A.}~\bibnamefont{Osipov}},
{\bibfnamefont{B.}~\bibnamefont{Hiller}}, 
{\bibfnamefont{J.}~\bibnamefont{Moreira}}, 
\bibnamefont{and} 
{\bibfnamefont{A.}~\bibfnamefont{H.}~\bibnamefont{Blin}},
\bibinfo{journal}{Eur. Phys. J.\ C} \textbf{\bibinfo{volume}{46}},
\bibinfo{pages}{225} (\bibinfo{year}{2006}); 
\bibinfo{author}{\bibfnamefont{A.}~\bibfnamefont{A.}~\bibnamefont{Osipov}},
{\bibfnamefont{B.}~\bibnamefont{Hiller}},
{\bibfnamefont{J.}~\bibnamefont{Moreira}},
{\bibfnamefont{A.}~\bibnamefont{H.}~\bibnamefont~{Blin}},
\bibnamefont{and} 
{\bibfnamefont{J.}~\bibnamefont{da Provid\^encia}},
\bibinfo{journal}{Phys. Lett.\ B} \textbf{\bibinfo{volume}{646}},
\bibinfo{pages}{91} (\bibinfo{year}{2007}); 
\bibinfo{author}{\bibfnamefont{A.}~\bibfnamefont{A.}~\bibnamefont{Osipov}},
\bibinfo{author}{\bibfnamefont{B.}~\bibnamefont{Hiller}},
\bibinfo{author}{\bibfnamefont{J.}~\bibnamefont{Moreira}},
\bibnamefont{and}
\bibinfo{author}{\bibfnamefont{A.}~\bibfnamefont{H.}~\bibnamefont{Blin}}, 
\bibinfo{journal}{Phys. Lett.\ B} \textbf{\bibinfo{volume}{659}},
\bibinfo{pages}{270} (\bibinfo{year}{2008});
\bibinfo{author}{\bibfnamefont{B.}~\bibnamefont{Hiller}},
\bibinfo{author}{\bibfnamefont{A.}~\bibfnamefont{A.}~\bibnamefont{Osipov}},
\bibinfo{author}{\bibfnamefont{A.}~\bibfnamefont{H.}~\bibnamefont{Blin}},
\bibnamefont{and} 
{\bibfnamefont{J.}~\bibnamefont{da Provid\^encia}},
\bibinfo{howpublished}{arXiv:hep-ph/0802.3193} (\bibinfo{year}{2008}). 

\bibitem[{\citenamefont{Boyd et al.}(1996)}]{Boyd}
\bibinfo{author}{\bibfnamefont{G.}~\bibnamefont{Boyd}},
\bibinfo{author}{\bibfnamefont{J.}~\bibnamefont{Engels}},
\bibinfo{author}{\bibfnamefont{F.}~\bibnamefont{Karsch}},
\bibinfo{author}{\bibfnamefont{E.}~\bibnamefont{Laermann}},
\bibinfo{author}{\bibfnamefont{C.}~\bibnamefont{Legeland}},
\bibinfo{author}{\bibfnamefont{M.}~\bibnamefont{L\"{u}tgemeier}},
\bibnamefont{and}
\bibinfo{author}{\bibfnamefont{B.}~\bibnamefont{Petersson}},
 \bibinfo{journal}{Nucl. Phys.} \textbf{\bibinfo{volume}{B469}},
\bibinfo{pages}{419} (\bibinfo{year}{1996}). 

\bibitem[{\citenamefont{Kaczmarek}(2002)}]{Kaczmarek}
\bibinfo{author}{\bibfnamefont{O.}~\bibnamefont{Kaczmarek}},
\bibinfo{author}{\bibfnamefont{F.}~\bibnamefont{Karsch}},
\bibinfo{author}{\bibfnamefont{P.}~\bibnamefont{Petreczky}},
\bibnamefont{and}
\bibinfo{author}{\bibfnamefont{F.}~\bibnamefont{Zantow}},  
  \bibinfo{journal}{Phys. Lett.\ B} \textbf{\bibinfo{volume}{543}},
  \bibinfo{pages}{41} (\bibinfo{year}{2002}).

\bibitem[{\citenamefont{Karsch}(2002)}]{Karsch3}
\bibinfo{author}{\bibfnamefont{F.}~\bibnamefont{Karsch}}, 
\bibinfo{journal}{Lect. notes Phys. } \textbf{\bibinfo{volume}{583}},
\bibinfo{pages}{209} (\bibinfo{year}{2002}). 

\bibitem[{\citenamefont{Karsch, Leermann and Peikert}(2002)}]{Karsch4}
\bibinfo{author}{\bibfnamefont{F.}~\bibnamefont{Karsch}}, 
\bibinfo{author}{\bibfnamefont{E.}~\bibnamefont{Laermann}}, 
\bibnamefont{and}
\bibinfo{author}{\bibfnamefont{A.}~\bibnamefont{Peikert}},
\bibinfo{journal}{Nucl. Phys. \ B} \textbf{\bibinfo{volume}{605}},
\bibinfo{pages}{579} (\bibinfo{year}{2002}). 

\bibitem[{\citenamefont{Cheng et al.}(2006)}]{MCheng}
\bibinfo{author}{\bibfnamefont{M.}~\bibnamefont{Cheng et al.}}, 
\bibinfo{journal}{Phys. Rev. \ D} \textbf{\bibinfo{volume}{74}},
\bibinfo{pages}{054507} (\bibinfo{year}{2006}). 

\bibitem[{\citenamefont{Weinberg}(1979)}]{Weinberg}
\bibinfo{author}{\bibfnamefont{S.}~\bibnamefont{Weinberg}}, 
\bibinfo{journal}{Physica } \textbf{\bibinfo{volume}{96A}}
\bibinfo{page} {327}(\bibinfo{year}{1979}); 
\bibinfo{author}{\bibfnamefont{S.}~\bibnamefont{Weinberg}}, 
\bibinfo{journal}{{\it "The quantum theory of fields II"}, Cambridge University Press}
(\bibinfo{year}{1996}). 


\bibitem[{\citenamefont{Gasser and Leutwyler}(1985)}]{Gasser}
\bibinfo{author}{\bibfnamefont{J.}~\bibnamefont{Gasser}}, 
\bibnamefont{and}
\bibinfo{author}{\bibfnamefont{H.}~\bibnamefont{Leutwyler}},
\bibinfo{journal}{Nucl. Phys. \ B} \textbf{\bibinfo{volume}{250}},
\bibinfo{pages}{465} (\bibinfo{year}{1985}); 
\bibinfo{author}{\bibfnamefont{J.}~\bibnamefont{Gasser}}, 
\bibnamefont{and}
\bibinfo{author}{\bibfnamefont{H.}~\bibnamefont{Leutwyler}},
\bibinfo{journal}{Phys. Rept. } \textbf{\bibinfo{volume}{87}},
\bibinfo{pages}{77}(\bibinfo{year}{1982}). 

\bibitem[{\citenamefont{DeGrand and DeTar}(2001)}]{DeGrand}
\bibinfo{author}{\bibfnamefont{T.}~\bibnamefont{A.}~\bibnamefont{DeGrand}}, 
\bibnamefont{and}
\bibinfo{author}{\bibfnamefont{C.}~\bibnamefont{DeTar}},
\bibinfo{journal}{Nucl. Phys. \ B} \textbf{\bibinfo{volume}{225}},
\bibinfo{pages}{590} (\bibinfo{year}{1983}). 

\bibitem[{\citenamefont{Patel}(2001)}]{Patel}
\bibinfo{author}{\bibfnamefont{A.}~\bibnamefont{Patel}}, 
\bibinfo{journal}{Nucl. Phys. \ B} \textbf{\bibinfo{volume}{243}},
\bibinfo{pages}{411} (\bibinfo{year}{1984}). 

\bibitem[{\citenamefont{Alford et al}(2001)}]{Alford}
\bibinfo{author}{\bibfnamefont{M.}~\bibnamefont{Alford}}, 
\bibinfo{author}{\bibfnamefont{S.}~\bibnamefont{Chandrasekharan}}, 
\bibinfo{author}{\bibfnamefont{J.}~\bibnamefont{Cox}}, 
\bibnamefont{and}
\bibinfo{author}{\bibfnamefont{U.}~\bibfnamefont{-J.}~\bibnamefont{Wiese}},
\bibinfo{journal}{Nucl. Phys. \ B} \textbf{\bibinfo{volume}{602}},
\bibinfo{pages}{61} (\bibinfo{year}{2001}). 

\bibitem[{\citenamefont{Kim}(2005)}]{Kim}
\bibinfo{author}{\bibfnamefont{S.}~\bibnamefont{Kim}}, 
\bibinfo{author}{\bibfnamefont{Ph.}~\bibfnamefont{de}~\bibnamefont{Forcrand}}, 
\bibinfo{author}{\bibfnamefont{S.}~\bibnamefont{Kratochvila}}, 
\bibnamefont{and}
\bibinfo{author}{\bibfnamefont{T.}~\bibnamefont{Takaishi}},
\bibinfo{howpublished}{arXiv:hep-lat/0510069} (\bibinfo{year}{2005}). 

\bibitem[{\citenamefont{Hasenfratz}(1983)}]{Hasenfratz}
\bibinfo{author}{\bibfnamefont{P.}~\bibnamefont{Hasenfratz}}, 
\bibnamefont{and}
\bibinfo{author}{\bibfnamefont{F.}~\bibnamefont{Karsch}},
\bibinfo{journal}{Phys. Lett.\ B} \textbf{\bibinfo{volume}{125}},
\bibinfo{pages}{308} (\bibinfo{year}{1983}). 

\end{thebibliography}
\end{document}